\documentclass[a4paper,floatfix,aps,prd,amsmath,amssymb,nofootinbib,preprintnumbers,superscriptaddress]{revtex4}

\usepackage{times,amsfonts}
\usepackage{natbib}
\usepackage[english]{babel}
\usepackage[T1]{fontenc}
\usepackage[latin2]{inputenc}
\usepackage{graphicx}
\usepackage{color}
\usepackage{tabularx}
\usepackage{mathrsfs}
\bibpunct{(}{)}{;}{a}{}{,}
\usepackage{aas_macros}

\usepackage{hyperref}
\usepackage{url} 

\newcommand\tabcaption{\def\@captype{table}\caption}
\makeatother

\newcommand\Nr{N_r}

\newcommand\chisq{\chi^2}

\newcommand{\lb}[2]{$(l,b)=(#1^\circ,#2^\circ)$}
\newcommand{\lmax}{\ell_{\rm{max}}}
\newcommand{\lmin}{\ell_{\rm{min}}}
\newcommand{\Alminlmax}{A_{\ell_{\rm min}}^{\ell_{\rm max}}}

\newcommand{\alm}{a_{\ell}^m}
\newcommand{\ClILCfit}{C_{\ell}^{\rm ILCfit}}
\newcommand{\Clfid}{C_{\ell}^{\rm fid}}
\newcommand{\alms}[1]{a_{\ell,{\rm #1}}^m}

\newcommand{\Plmrecurrence}{(\ell-m+1) P_{\ell+1}^m(x) = (2 \ell+1) x P_\ell^m(x) - (\ell+m) P_{\ell-1}^m (x)}
\newcommand{\YlmrecurrenceRHS}[1]{\frac{\ell-m+1}{2 \ell+1} Y_{\ell+1}^m(#1) + \frac{\ell+m}{2 \ell+1} Y_{\ell-1}^m(#1)}

\begin{document}
\title{Hemispherical power asymmetry: parameter estimation from CMB WMAP5 data.}
\author{Bartosz Lew} \email[]{blew@a.phys.nagoya-u.ac.jp }
\affiliation{Department of Physics and Astrophysics, Nagoya University, Nagoya 464-8602, Japan}
\affiliation{Toru\'n Centre for Astronomy, Nicolaus Copernicus University, ul. Gagarina 11, 87-100 Toru\'n, Poland}

\begin{abstract}

We re-examine the evidence of hemispherical power asymmetry, detected in the cosmic microwave background (CMB) WMAP 
(Wilkinson Microwave Anisotropy Probe) data using a new method.
We use a data filtering, preprocessing , and a statistical approach different from those used previously, and 
pursue an independent method of parameter estimation.
First, we analyze the hemispherical variance ratios and compare these with  simulated distributions.
Secondly, working within a previously proposed CMB bipolar modulation model,
we constrain the model parameters: the amplitude and the orientation of the modulation field
as a function of various multipole bins.
Finally, we select three ranges of multipoles leading to the most anomalous signals, 
and we process corresponding 100 Gaussian, random field (GRF) simulations, treated as observational data, 
to further test the 
statistical significance and robustness of the hemispherical power asymmetry.
For our analysis we use the Internally-Linearly-Coadded (ILC) full sky map, and the KQ75 cut sky V channel 
foregrounds reduced map of the WMAP five year data (V5).
We constrain the modulation parameters using a generic maximum a \emph{posteriori} method.

In particular, we find differences in hemispherical power distribution, which when described in terms of
a model with bipolar modulation field, exclude the field amplitude value of the isotropic model $A=0$ 
at confidence level of $\sim 99.5\%$ ( $\sim 99.4\%$) in the multipole range $\ell\in[7,19]$ 
($\ell\in[7,79]$) in the V5 data, 
and at the confidence level $\sim 99.9\%$ in the multipole range $\ell\in[7,39]$ in the ILC5 data, with
the best fit (modal PDF) values in these particular multipole ranges of $A=0.21$ ($A=0.21$) and $A=0.15$ respectively.

However, we also point out that similar or larger significances (in terms of rejecting the isotropic model), 
and large best-fit modulation amplitudes are obtained
in GRF simulations as well, which reduces the overall significance of the CMB power asymmetry down to only about 
$94\%$ ($95\%$) in the V5 data, in the range $\ell\in[7,19]$ ($\ell\in[7,79]$). 
\end{abstract}

\maketitle

\section{Introduction}

The Gaussianity and the statistical isotropy of the fluctuations in the Cosmic Microwave Backgrounds Radiation (CMBR)
are two generic features of current standard cosmological model and are compatible with the simplest inflationary scenarios.
These predictions have been extensively studied in number of works. An incomplete list includes:
\cite{komatsu-2003-148,2008arXiv0803.0547K,2006MNRAS.371L..50M,2004ApJ...609...22V,wiaux-2006-96,2004ApJ...613...51M,
2004MNRAS.349..973S,2007arXiv0712.1118N,2004PhRvD..69f3007C,2006MNRAS.tmp..497C,2005MNRAS.358..684C,2006astro.ph..6394C,
2008arXiv0804.0136C,2007A&A...474...23C,2007NewAR..51..250D,2008arXiv0807.2687A,2006astro.ph..7577S,2006PhRvD..74l3521H,2006PhRvD..74l3521H,2008MNRAS.385.1718S,2004MNRAS.354..641H,2004ApJ...607L..67H,PASHtestbyBernui,2007IJMPD..16..411B,2005PhRvD..72f3512N,2003ApJ...590L..65C,2006ApJ...647L..87C,2003MNRAS.346...47G,WMAPmultipol,2006MNRAS.367...79C,2006astro.ph..5135C,2005ApJ...635..750B,2006PhRvD..74f3506A,2006PhRvD..74b3005D,2005PhRvL..95g1301L,2007MNRAS.378..153L,2005ApJ...629L...1J,EfstNoProb03a,2004ApJ...605...14E,2007ApJ...660L..81E,2008ApJ...672L..87E,2005PhRvL..95g1301L,2006astro.ph..8129P,2002MNRAS.331..865S,2001PhRvL..87y1303W,2004MNRAS.349..313P,LR08dodec,Lew2008,2005PhRvD..71d3002D,2008PhRvL.100r1301Y} and references
therein. 
Within the theory of inflation the primordial fluctuations are expected
to form a Gaussian Random Field (GRF) at the leading order in perturbation theory.
Their statistical properties are imprinted in the CMB fluctuations, providing an interesting window
on the processes of the early Universe. Although the instrumental effects, like
non-Gaussian and non-isotropic noise, or eccentric beams, and 
astrophysical foregrounds effects, like Galactic, and extra-Galactic point sources, and extended sources of emission, 
are either well controlled, or corrected for, or masked out, 
a set of an unexpected anomalies of various magnitudes and at various scales have been detected in the current CMB data 
\citep{2007ApJ...655...11C,2007IJMPD..16..411B,2005MNRAS.357..994L,WMAPmultipol,2006MNRAS.367...79C,2006PhRvD..74f3506A,2007MNRAS.378..153L,2004ApJ...605...14E,2005PhRvL..95g1301L,2005PhRvD..72j1302L,2006PhRvD..74h3509C,2008arXiv0804.2387D} (see also 
\cite{2006NewAR..50..868H,2008arXiv0805.4157M,2008arXiv0807.1816C} for recent reviews and references therein).
These anomalies call for plausible theoretical 
explanations since, if robustly detected, these can be used as valuable observables of the physics of the early
Universe, or a new window on some of the late time effects 
\citep{2007ApJ...664..650I,2008arXiv0806.0377E,2008MNRAS.tmp..929B,2007arXiv0710.5897A,2008arXiv0804.2387D,2005PhRvD..72f3002B,2008PhRvD..77f3008D}.

In this paper we re-investigate the well-known, hemispherical power asymmetry observed in the CMB maps.
We revisit the properties of the asymmetry, constrain parameters of the previously proposed bipolar modulation field model
\citep{2005PhRvD..72j3002G} responsible for generation of the asymmetry, 
and we estimate the statistical significance using a generic maximum likelihood method, 
and realistic Monte-Carlo simulations.
Given that we introduce and utilize a different method, from those previously used, and rely on different
assumptions, while relax some other, our results can serve as a separate cross consistency and stability check.
We provide a through tests of the method so as to validate the presented results.

For the first time we estimate the parameters of the hemispherical modulation for selected ranges of multipoles. 
Finally, we assess the significance of the hemispherical power asymmetry via direct comparison 
to the GRF CMB simulations.

The main differences from, and extensions to the previous analyses are: 
(i) we rely on the local  real-space measurements of the variance as an estimator for the power asymmetry, 
(ii) we do not assume any priors on the probability distribution function (PDF) for any of the modulation parameters, and 
explore the likelihood function in the full (albeit sparse) parameter space and apply interpolation, 
(iii) we fully include the effects of the cut-sky, cross-multipoles power leakage and 
(iv) we analyze the power asymmetry in the various slices through the spherical harmonic space, pre-filtering the data
 prior to the analysis, rather than considering all scales scrambled together.

Since the full exploration of the parameter space is CPU expensive, our analysis is based on a few  
assumptions that greatly simplify, and speed-up the parameter estimation process. Using this different 
approach, while providing tests and justification for the 
assumptions made, we give a new estimates on the significance of the hemispherical power asymmetry anomaly.

We also discuss limitations of usage of the method with regard to the extent to which the assumptions of the  
method remain acceptably valid.

The organization of the paper is as follows:
In Section ~\ref{sec:data_and_sims} we describe our datasets and CMB simulations.
In Section ~\ref{sec:power_ratio} we present results of a statistics that measure the hemispherical power ratios.
In Section ~\ref{sec:modulation_parameter_estimation} we focus on the properties of 
the power modulation model, our assumptions and tests of the assumptions,
and then detail on our method for modulation parameter estimation.
In Section ~\ref{sec:tests} we present the results of various tests of the method.
Results of the application of the method to the real CMB data are presented in section ~\ref{sec:results}.
Discussion and conclusions are given in sections~\ref{sec:discussion} and~\ref{sec:conclusions} respectively.

\section{Data and simulations}
\label{sec:data_and_sims}

For the main analysis in the paper we use the WMAP five-year foreground reduced CMB temperature maps 
\citep{2008arXiv0803.0732H} 
from differential assemblies (DA) V1 and  V2, because these spectral channels provide the best trade off between 
foregrounds of different spectral properties (i.e. the blue tilted galactic dust emission, 
and red-tilted galactic synchrotron and free-free emissions).
We co-add these observations (and corresponding simulations) using the inverse noise pixel weighting scheme. 
We will refer to the resulting map as V5.
We generate simulations using the fiducial best fit $\Lambda CDM$ model power spectrum of \cite{2008arXiv0803.0586D} 
(constructed using the mean likelihood parameters) which we call $\Clfid$.

Also, for comparison purposes, we will use the five year release of the Internally-Linearly-Coadded (ILC5) 
map and also for additional tests the Harmonic-Internal-Linear-Combination (HILC5) map \citep{2008arXiv0803.1394K}.

\begin{figure}[!t]
\centering
\includegraphics[angle=-90,width=0.6\textwidth]{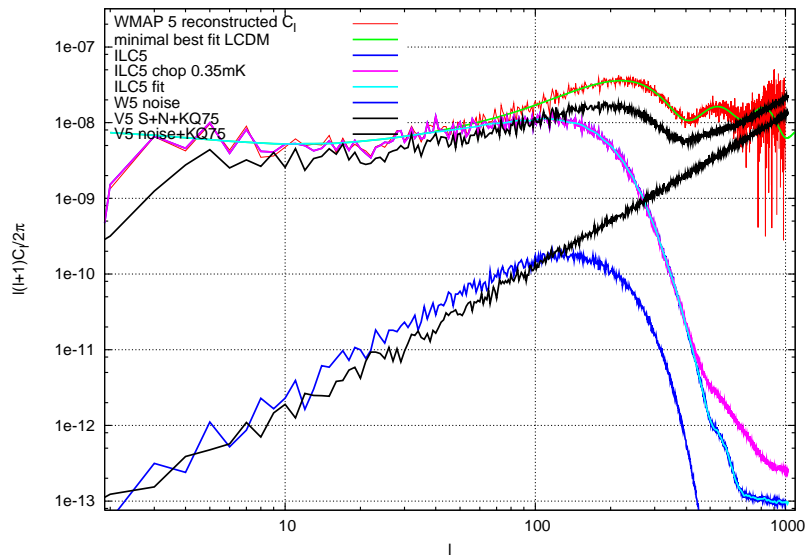}
\caption{Power spectra of the data sets used in the analysis. 
We generate the GRF simulations of the V5 data using the best fit minimal $\Lambda$CDM model (green line)
to the reconstructed WMAP5 data power spectrum (red line).
The corresponding cut-sky pseudo power spectrum from one of the V5 simulations and V5 noise power spectrum are
plotted in black (the top-black curve and the bottom-black curve respectively). 
Simulations of the ILC5 data are created using the fitted power spectrum $\ClILCfit$ (light blue). 
The reconstructed from the ILC5 data power spectrum is plotted using the dark blue (top) line. 
The noise power of the noisiest channel of the WMAP data (W) smoothed with one-degree (FWHM) Gaussian beam
is also plotted (dark-blue bottom) and used here to place an upper limit constraint on the amount of 
noise in the ILC data. 
The power spectrum of the ILC5 data chopped at temperature threshold of $350\mu {\rm K}$ is also shown (pink line). 
In the parameter estimation analysis we will use only the range of multipoles $\ell\leq 80$.
}
\label{fig:data_powers}
\end{figure}

In Fig.~\ref{fig:data_powers} we plot the power spectra of the data sets that will be used in the power 
modulation parameters estimation analysis. 
For the purpose of the analysis (to be explained latter) we create a fitted power spectrum to
the ILC5 data by concatenating
the $\Clfid$ power spectrum in the limit of low multipoles ($\ell<30$), with
the cubic spline fit to the piece-wise averaged, full-sky power spectrum, reconstructed from the ILC5 data for 
multipoles $\ell \geq 30$. We will call this fitted spectrum $\ClILCfit$.
We cut off the residual strong foregrounds in the ILC5 map at the $\pm 350\mu {\rm K}$ threshold; a level estimated from
GRF foregrounds-free simulations.

Note that in the analysis of the modulation parameter estimation we will focus only on the large scale multipoles 
($\ell\leq 80$), where the differences in the power spectrum due to the $350\mu {\rm K}$ chop of the residual 
galactic contamination remaining the ILC5 data are completely unimportant, 
and where the signal-to-noise (SNR) ratio yields $\rm{SNR} \gtrsim 10^2$ (Fig.~\ref{fig:data_powers}).

Throughout the paper we will use the V5 data along with the KQ75 sky mask unless otherwise noted.
For comparison purposes, the ILC5 and HILC5 datasets are used without any masks throughout of the paper
\footnote{Although we realize that the usage of the unmasked ILC maps can lead to somewhat overestimated
constraints on any power asymmetry signals due to residual foregrounds contamination, we include the 
analysis of these maps in the full considered multipole range ($\ell \leq 80$) mostly for comparison purposes.}.

\section{Hemispherical power ratio}
\label{sec:power_ratio}
We begin the analysis of the hemispherical power asymmetry by computing the following statistics:
\begin{equation}
\label{eq:sNsS}
\begin{array}{lll}
R_{\rm NS,\lmax} &=& \max\limits_{\mathbf{ \hat n_s}} \Bigl( \frac{\sigma_{\rm N}(\mathbf{ \hat n_s},\lmax)}{\sigma_{\rm S}(\mathbf{ \hat n_s},\lmax)}  \Bigr)\\\\
r_{\rm NS,\lmax} &=& \min\limits_{\mathbf{ \hat n_s}} \Bigl( \frac{\sigma_{\rm N}(\mathbf{ \hat n_s},\lmax)}{\sigma_{\rm S}(\mathbf{ \hat n_s},\lmax)}  \Bigr)\\\\
\end{array}
\end{equation}
These are simply the maximal ($R_{\rm NS,\lmax}$) and the minimal ($r_{\rm NS,\lmax}$) ratios of the hemispherical 
standard deviations, 
found in the all sky search over directions $\mathbf{ \hat n_s}$, that define the orientation of two hemispheres. 
The $\sigma_{\rm N}$ and $\sigma_{\rm S}$ values are the cut-sky (in case of V5 data) 
or full-sky (in case of ILC5/HILC5 data) hemispherical standard deviations of a map.

To define the grid of directions ($\mathbf{ \hat n_s}$), 
we choose to use the first 96, ring-ordered directions, defined by the 
pixel centres of the HEALPIX pixelization scheme \citep{Healpix2005} of resolution parameter $n_s=4$.
The measurements are performed using V5, ILC5 and HILC5 datasets within either, chosen ranges of multipoles, or
as a function of a cumulative maximal multipole number $\lmax$.

\begin{figure}[!ht]
\centering
\begin{tabular}{c}
V5 data\\
\includegraphics[height=0.22\textheight,width=0.8\textwidth]{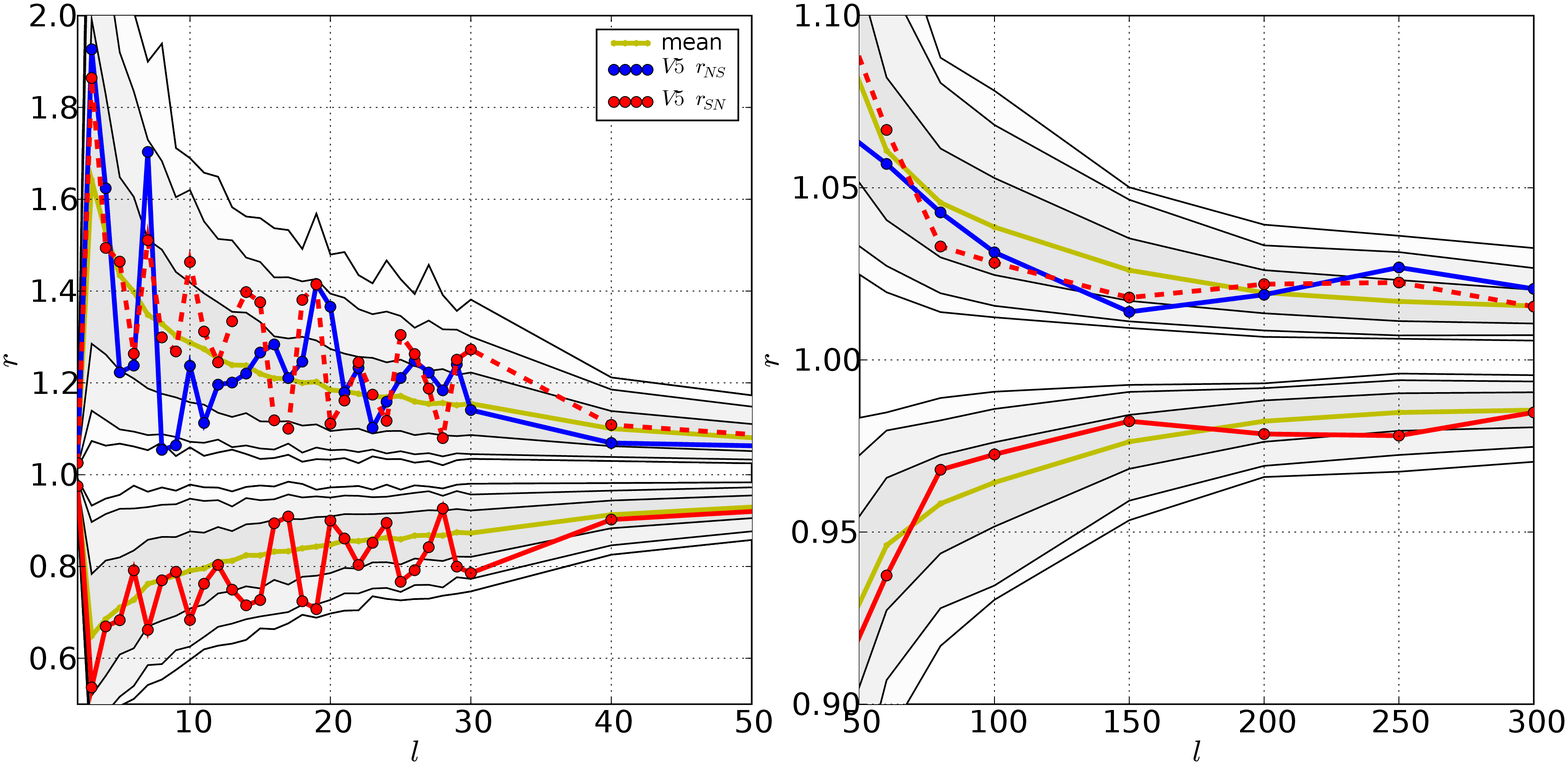}\\
\includegraphics[height=0.22\textheight,width=0.8\textwidth]{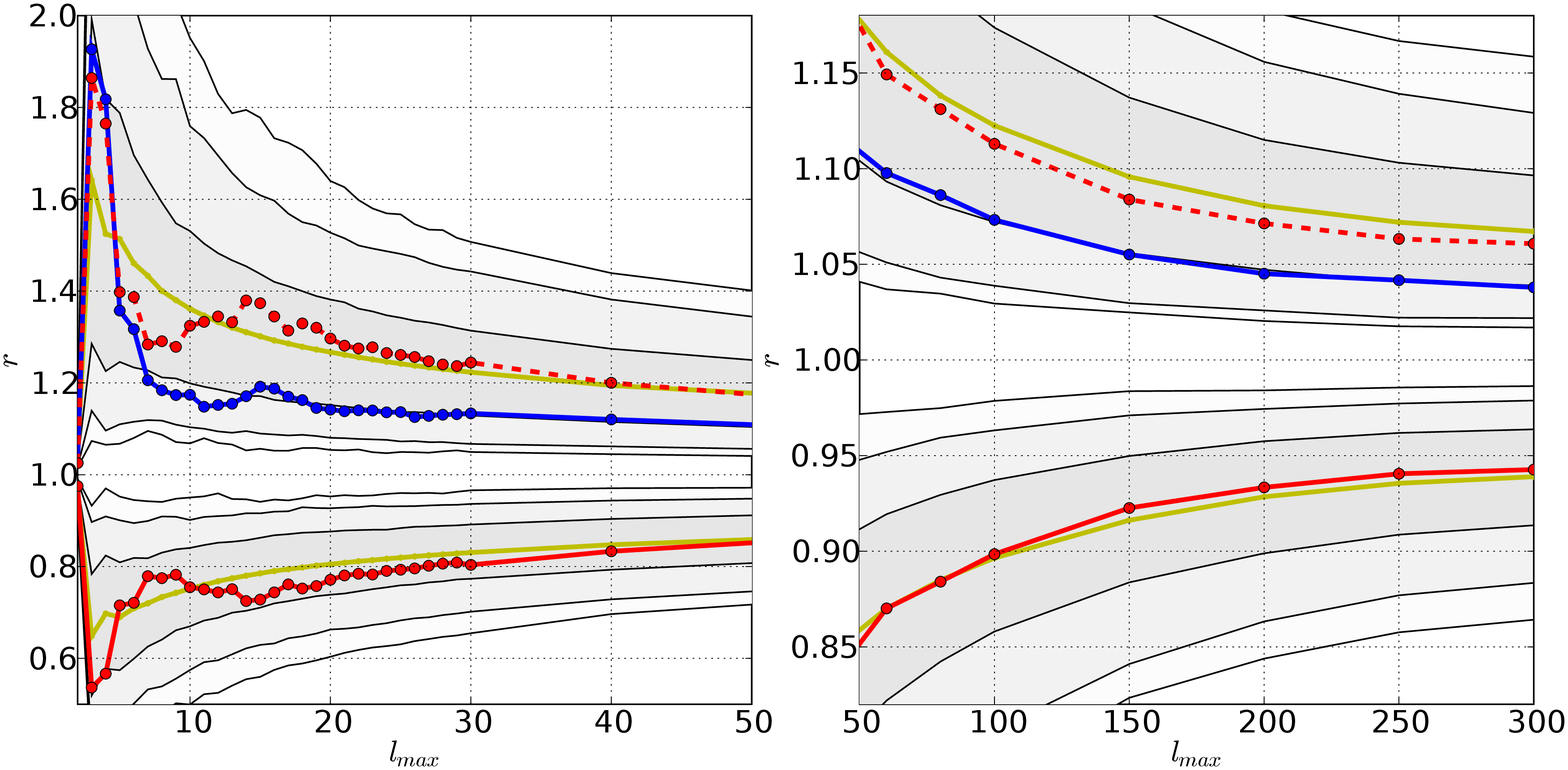}\\
ILC5 \& HILC5 data\\
\includegraphics[height=0.22\textheight,width=0.8\textwidth]{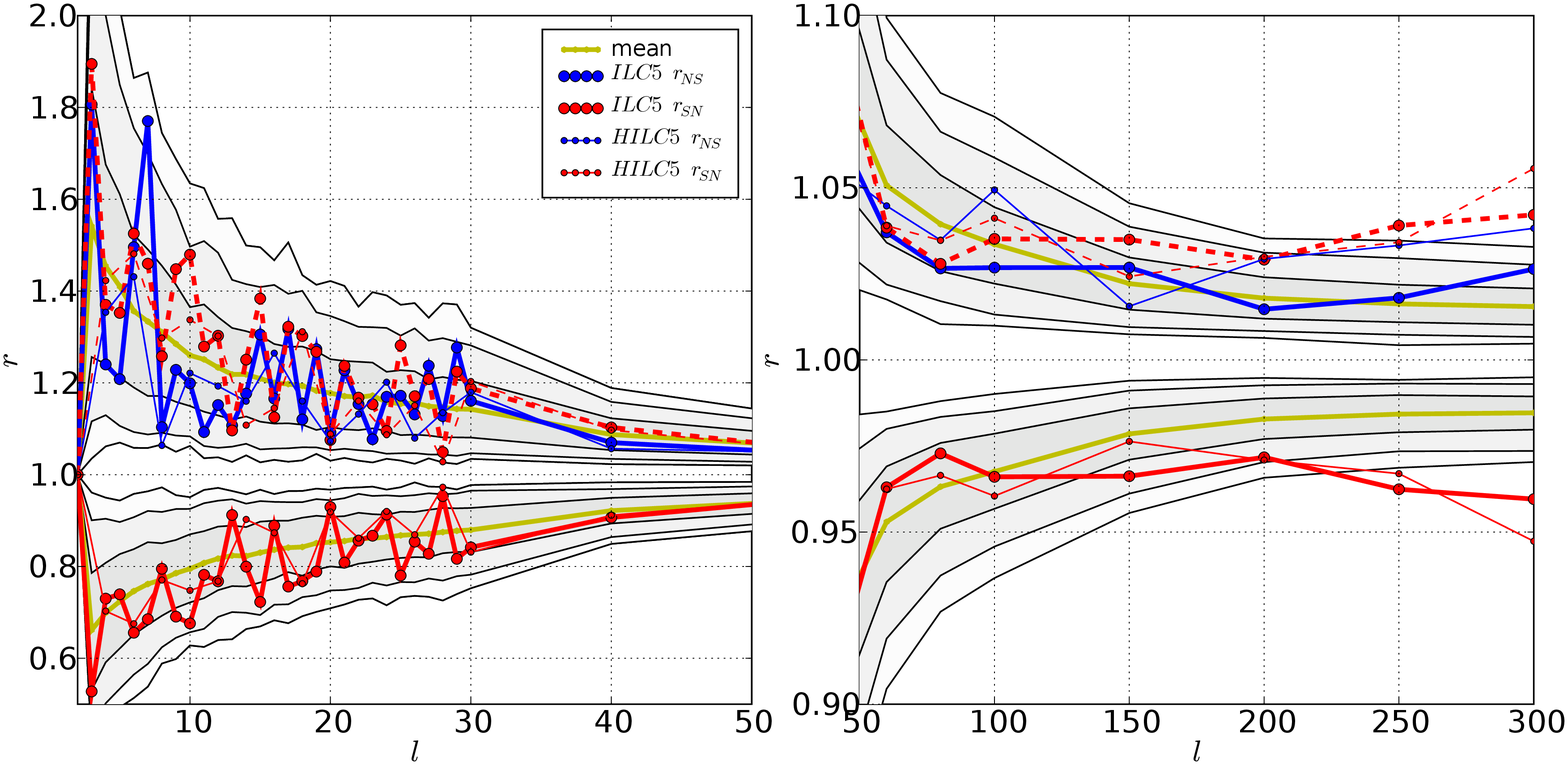}\\
\includegraphics[height=0.22\textheight,width=0.8\textwidth]{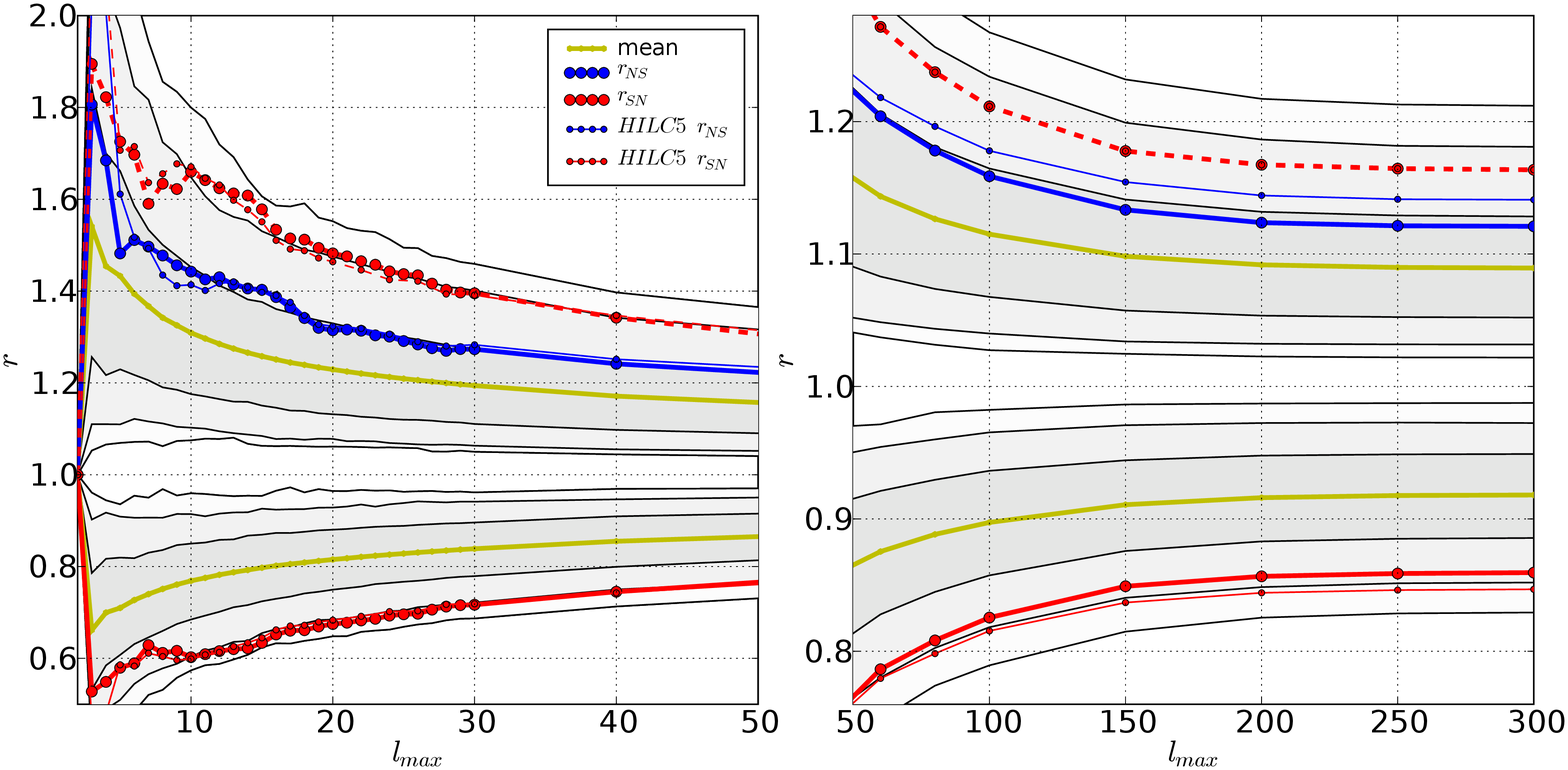}
\end{tabular}
\caption{Hemispherical power asymmetry in filtered, in the spherical harmonic space, V5 data (top panel) and
in the ILC5 and HILC5 data (bottom panel).
In the first row of each panel we plot the maximal ($R_{\rm NS,\lmax}$) (blue)/minimal ($r_{\rm NS,\lmax}$) (red) 
ratios of the standard deviations
as a function of the filtered range of multipoles defined by the two $\ell$ values corresponding to any of the two 
neighboring plotted points (see text for details).
In the second row of each panel we plot the ratios as a function of the cumulative maximal multipole number $\lmax$ 
defining the considered range of filtered multipoles: $\ell \in [2,\lmax]$.
The red (dashed) line is the inverted minimal ($1/r_{\rm NS,\lmax} = r_{\rm SN,\lmax}$) 
ratio of the hemispherical standard deviations, plotted to facilitate appreciation of the power asymmetry by 
direct comparison with the blue lines.
The gray bands represent the 68\%, 95\% and 99\% confidence level contours. 
The simulations' mean is plotted with yellow solid line.
The hemispherical asymmetry mostly appears to be confined to the multipoles range $\ell_{\rm assym}\in[8,15]$ in V5, ILC5 
and HILC5 dataset. 
}
\label{fig:sigmaNS}
\end{figure}

The results of this survey are summarized in Fig.~\ref{fig:sigmaNS}

The hemispherical asymmetry in the filtered bins of multipoles seems to be localized within the range of multipoles 
\mbox{$\ell_{\rm assym}\in[8,15]$} in all of the dataset: V5, ILC5 and HILC5.

Note that the first point in every plot: i.e. the quadrupole ($\ell=2$), has the ratios always equal unity, 
because only the quadrupole map is used in variance measurements, and since the single-multipole maps have a 
point (anty)symmetry, due to the properties of the spherical harmonics, the variance is identical in the 
two hemispheres. 
In the range of multipoles $\ell\in (2,30)$ we process the
maps containing only two neighboring multipoles: eg. for the $\ell=3$ we use combined maps of multipoles 
$\ell=2$ and  $\ell=3$; for $\ell=4$ we use combined maps of multipoles $\ell=3$ and $\ell=4$, and so on. 
For higher multipoles ($\ell>30$) the bins are larger, and are defined by the two neighboring plotted points in Fig.~\ref{fig:sigmaNS}.
In the case of the cumulative plots (second and fourth row in Fig.~\ref{fig:sigmaNS}) 
we use all multipoles from $\ell=2$ up to $\lmax$,
and due to the cumulative process, the curves on the right-hand side plots, 
(showing the statistics in the large-$\ell$ limit), do not exhibit much of details,
since most of the map power (variance) comes from the low multipoles. 
In the limit of large multipoles, the multipole range-filtered maps 
show much more details as they are not overwhelmed by the power of the low multipoles. 
In particular, notice the strong, systematical
deviation away from the simulation mean in case of the ILC5 and HILC5 data, starting with multipoles $\ell \gtrsim 150$.
These are most likely caused by the extended foregrounds, and point sources remaining in the maps, 
since we do not apply any sky cuts with these data.
We over-plot the results for the HILC5 data using the confidence level contours derived from the ILC5 simulations,
just to compare them with the results obtained for  the ILC5 data.

We also notice that  in case of the multipole range-filtered V5 data, 
(top row in Fig.~\ref{fig:sigmaNS}) some asymmetry is also seen in the range $\ell\in[29,40]$.
In the same plot it appears that the ``northern'' hemisphere is anomalous due to the decrement in power
as compared to the simulations in multipole range \mbox{$\ell_{\rm assym}\in[8,15]$}.

For the case of the multipole, range-filtered results (first and third row in Fig.~\ref{fig:sigmaNS}) 
we derive the joint ``probability of  rejecting'', using a generic multivariate calculus,
and for the V5 data we obtain result consistent with the simulations at $< 82\%$  confidence level for both
the maximal hemispherical ratio ($R_{\rm NS,\lmax}$) and the minimal hemispherical ratio ($r_{\rm NS,\lmax}$) 
regardless of whether or not the off-diagonal terms of the covariance matrix are included.
This result corresponds to the full range of multipoles up to $\ell =300$.

As a final note to this analysis we point out, that the low significance of the power asymmetry, illustrated
in Fig.~\ref{fig:sigmaNS}, may result from the fact that we analyzed the data in a thin slices through the 
spherical harmonic space (every two multipoles) up to $\ell = 30$.
Although the full covariance matrix analysis should in principle, be stable to that, it's possible 
that other binning of the data may lead to somewhat different result.
In order to check that, we similarly estimated the joint ``probability of rejecting'' within the multipole range
\mbox{$\ell_{\rm assym}\in[8,15]$} for the V5 data, but we found the data to be consistent with simulations
at confidence level as low as $60\%$.

In the following we will constrain the properties of the hemispherical power asymmetry in greater detail.

\section{Modulation parameters estimation}
\label{sec:modulation_parameter_estimation}

In \cite{2007ApJ...660L..81E} an approach for estimating the bipolar modulation parameters was concisely outlined, 
and was implemented to obtain the constraints on the modulation parameters.
In that work a Gaussian PDF form for all model parameters was assumed, except for the modulation orientation 
axis, for which a flat PDF form was used. As will be shown, the exact shape of the likelihood function may
have and important effect on the overall significance of the power asymmetry anomaly, so in contrast to that work
we directly reconstruct the likelihood function using a grid method. 

Since the full parameter space operations are time consuming, we introduce a few assumptions that greatly simplify 
the parameter estimation process. In the next section we discuss them one by one, and provide appropriate justifications.

\subsection{Bipolar modulation model parametrization}

\label{sec:parametrization}

We generalize the parametrization of the CMB modulation, form the one defined in our previous work \citep{Lew2008}, 
to account for a modulation that is effective only within a requested range of multipoles $(\lmin,\lmax )$. 
A CMB observation $T(\mathbf{\hat n})$ of the GRF CMB realization 
($T_{CMB}(\mathbf{\hat n})$) within a bipolar modulation model can be written as:
\begin{equation}
T(\mathbf{\hat n}) = \mathbf{B}(\mathbf{\hat n},\mathbf{\hat n'}) \star \Bigl( T_{CMB}(\mathbf{\hat n'}) \bigl(1+ M(\mathbf{\hat n'})\bigr) +  F(\mathbf{\hat n'}) \Bigr) w(\mathbf{\hat n}) +  N(\mathbf{\hat n}) w(\mathbf{\hat n})
\label{eq:observation}
\end{equation}
where the modulation field $M(\mathbf{\hat n})$ is defined as:
\begin{equation}
M(\mathbf{\hat n}) = \Alminlmax  \mathbf{\hat m} \cdot \mathbf{\hat n}
\label{eq:modulation}
\end{equation}
where $\mathbf{\hat n}$ is a unit vector and $M$ is a bipolar modulation field,
oriented along direction $\mathbf{\hat m}$ with amplitude 
$\Alminlmax$, which modulates the CMB component only within the specified range of multipoles between $\ell_{\rm min}$ and
$\ell_{\rm max}$. The $F(\mathbf{\hat n'})$ and $N(\mathbf{\hat n'})$ denote the residual foregrounds and the noise component respectively.
The $\mathbf{B}(\mathbf{\hat n},\mathbf{\hat n'})$ represents the real-space beam convolution kernel of the instrument, or any other effective convolution 
that has been applied to the data. The $w(\mathbf{\hat n})$ is the mask window function which can assume either $0$ for masked pixels or $1$ for unmasked pixels.
In case of the ILC5 data $w = 1$.
We will constrain the parameters $\Alminlmax$ and $\mathbf{\hat m}$ in different ranges of multipoles in order to  
investigate the modulation as a function of scale.

\subsection{Assumptions}

\label{sec:assumptions}
To facilitate the  reconstruction of the  multidimensional likelihood function in the following analysis we will rely on three assumptions.
We assume that 
(i) the noise in our dataset in the range of multipoles under consideration can be neglected, 
(ii) that the dataset maps are foregrounds free, and 
(iii) that the residual systematical effects of the modulation-induced change to the underlying power spectrum of the CMB does not
significantly influences the modulation parameter estimates.
In the following sections we will discuss each of the assumptions in greater detail.

\subsubsection{Signal to noise ratio}
\begin{figure}[!t]
\centering
\includegraphics[width=0.49\textwidth]{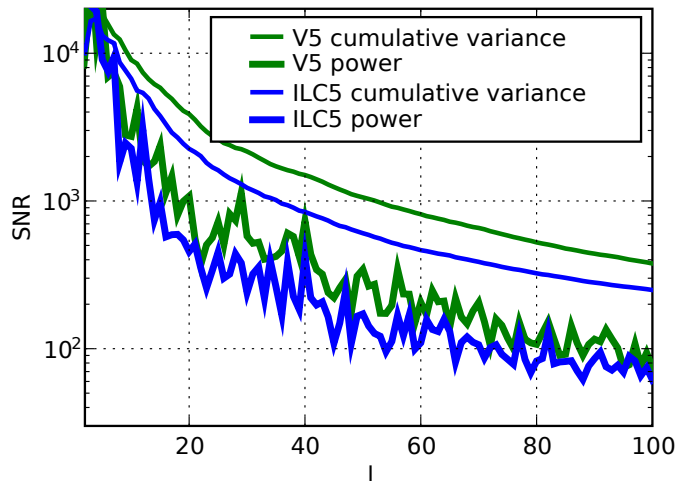}
\caption{An estimate of the signal-to-noise ratio (SNR) per multipole in the pseudo-power spectrum 
($C_\ell^{\rm signal}/C_\ell^{\rm noise}$) (thick lines) in the ILC5 (blue) and V5 (green) dataset
and the signal-to-noise ratio of the cumulative variance measurements (thin lines) defined as: 
\mbox{${\rm SNR}(\ell)=\sum_{j=2}^\ell(2 j+1)C_j^{\rm signal}/\sum_{j'=2}^\ell(2 j'+1)C_{j'}^{\rm noise}$}
in the ILC5 (blue) and V5 (green) dataset.}
\label{fig:snr}
\end{figure}

For the purpose of the analysis we assume that the noise in the CMB dataset (or simulations) 
has no significant impact on the CMB component modulation parameters estimates.
The accuracy of this assumption is scale dependent.
In Fig.~\ref{fig:snr} we plot the signal to noise ratio (SNR) as deduced from Fig.~\ref{fig:data_powers}.

We choose to work with the filtered data at scales below $\ell\leq 80$, where the signal-to-noise ratio 
per single multipole is approximately $SNR\sim 10^2$ (Fig.~\ref{fig:snr} thick lines).
Note that for the case of a real-space, variance measurements in a map composed of a range of multipoles, 
the SNR is larger due to the cumulative effect (Fig.~\ref{fig:snr} thin lines).

\subsubsection{Foregrounds}
We rely on the foregrounds reduced data of the cleanest WMAP5 channel - V  and a conservative sky mask KQ75. 
As for the ILC5 data the residual galactic foregrounds are clearly seen in the map. 
We drastically reduce their amplitude by a sharp cut at the level of $\pm 350 \mu {\rm K}$ 
(limit deduced from foregrounds free simulations). Of course this doesn't remove
the foregrounds but somewhat reduces their impact on the regional variance measurements. 
It should be noted that due to the residual foregrounds, a caution is recommended by the WMAP team
when analysing this map at scales $\ell \gtrsim 10$ \citep{2007ApJS..170..288H}. 
We present the results of the full sky ILC5 analysis for comparison purposes 
with the results obtained using an  extensively masked V5 data.
The V5 data should be therefore  more reliable in the limit of large multipoles.
However it will be interesting to compare the results between the two analyses both in the limits of low multipoles,
where the ILC map should be reliable, and in the limit of large multipoles, where some residual foreground contaminations
are present.

\subsubsection{Modulation effects to the power spectrum}
\label{sec:modulated_power_spectrum}
The modulation inevitably leads to a change in the underlying power spectrum at all scales, due to the 
multiplicative dipole component. Assuming a modulation orientation along the ``z-axis'' direction, 
the modulation field $M$  expressed by a spherical harmonic of degree 1 and order 0 is:
$M(\mathbf{\hat n}) = \Alminlmax a_1^0 Y_1^0(\mathbf{\hat n}) = \Alminlmax a_1^0 \cos(\theta)$,
\footnote{We use the \cite{AbramovitzandStegunbook} notation convention, and their phase definition 
of the spherical harmonics throughout the paper.}
where $a_1^0 = 2\sqrt{\frac{\pi}{3}}$,
and the spherical harmonic expansion of the modulated map reads:
\begin{equation}
T(\mathbf{\hat n}) = \sum_{\ell,m} a_\ell^m Y_\ell^m(\mathbf{\hat n}) + 2\sqrt{\frac{\pi}{3}} \Alminlmax \sum_{\ell,m} a_\ell^m \cos(\theta) Y_\ell^m(\mathbf{\hat n})
\label{eq:SHdecomp}
\end{equation}
where the first term corresponds to the initial CMB component and the second term corresponds to the modulated component.

Using the recurrence formula for the associated Legendre polynomials
\begin{equation}
\label{eq:Plmrecurrence}
\Plmrecurrence
\end{equation}
it is straightforward to see that
\begin{equation}
\label{eq:Ylmrecurrence}
\cos(\theta) Y_\ell^m(\mathbf{\hat n}) = \YlmrecurrenceRHS{\mathbf{\hat n}}
\end{equation}
where $x\equiv \cos(\theta)$.

According to Eq.~\ref{eq:Ylmrecurrence} the modulation leads to redistribution of power of a given multipole $\ell$ 
of the modulation component map on to 
the two neighboring multipoles $\ell -1$ and $\ell +1$ totally removing power from the multipole $\ell$.
Of course the power in the $\ell$th multipole is restored by redistribution of power of the $\ell+1$ and $\ell-1$  multipoles.
In general, the power is redistributed not only within the same $m$ mode but also within the $m\pm 1$ modes
when the modulation orientation is allowed to assume an arbitrary orientation.

\begin{figure}[!t]
\centering
\includegraphics[width=0.49\textwidth]{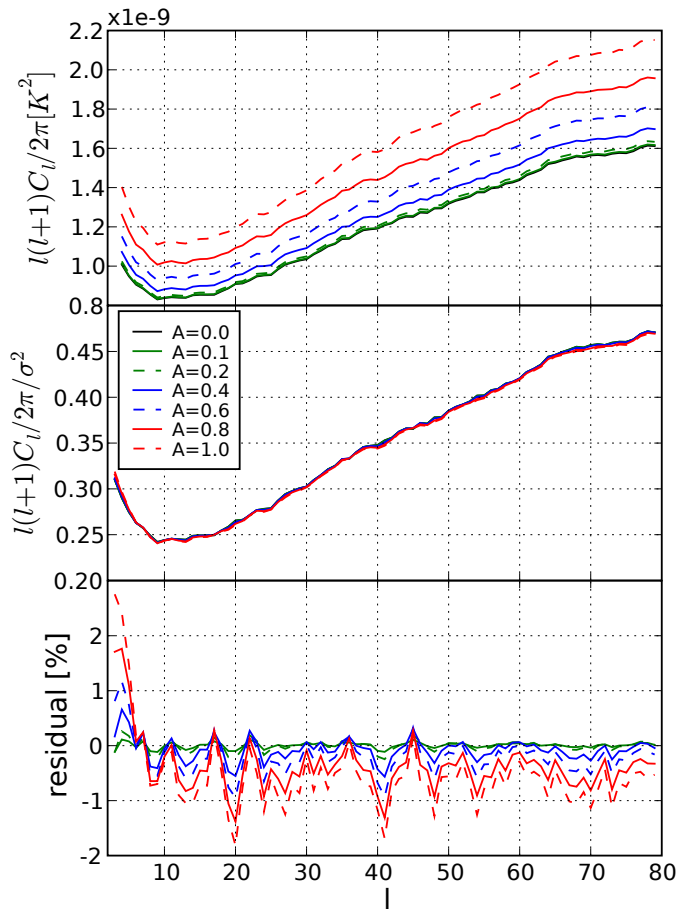}
\caption{{\it Top panel:} modulation-induced systematical changes to the fiducial power spectrum as inferred from the  average power spectrum 
from a sample of 1000 simulations. The overall shift, approximately by a constant factor in all multipoles is clearly seen.
{\it Middle panel:} average power spectra of the modulated maps, calibrated in real space, by the standard deviation.
{\it Bottom panel:} residual, fractional systematical deviation from the original (non-modulated) power spectrum.
}
\label{fig:modulation_power}
\end{figure}

It can also be shown that the modulated map has statistically more power than the corresponding non-modulated map.
This is clearly seen in the modulated map power spectrum (Fig.~\ref{fig:modulation_power} top panel) as a systematic 
departure from the initial power spectrum, whose
magnitude depends on the modulation amplitude $\Alminlmax$. 
In our analysis we will account for this systematical effect by means of a calibration
of the map in the real space by the standard deviation (Fig.~\ref{fig:modulation_power} middle panel). 
This approach does not eliminates the systematical effects, however it reduces them 
to a sub-percent levels (Fig.~\ref{fig:modulation_power} bottom panel) for the modulations that are
cosmologically relevant.

We study these systematical effects using a sample of 1\,000 full sky ILC5 simulations, which we modulate to various 
extent. Then we reconstruct the average modulated power spectra, and compare it to the average non-modulated 
power spectra (Fig.~\ref{fig:modulation_power}).
Note that e.g. for modulations $A\lesssim 0.6$ the systematical effects of the modulation are smaller than $1\%$ 
after a proper re-calibration of the map.
In practice the larger modulations  will be much stronger penalized due to the violation of the statistical isotropy,
than due to the systematical deviations from the initial, fiducial power spectrum.
It is also worth noting that the deviation from the initial power spectrum in the case of smaller, 
and much more relevant modulations is negligible, after variance recalibration 
(green curves in the bottom plot of Fig.~\ref{fig:modulation_power}).

Since we will be working with selected, filtered ranges of multipoles (in order to investigate the modulation hypothesis
as a function of scale) the multipoles at the upper and lower ends of that range will have their power significantly
suppressed, due to the modulation-induced power-leakage outside the chosen range. To improve the effectiveness of the calibration 
process, for a given multipole range of interest $[\ell_1,\ell_2]$ we will calibrate the modulation-altered simulations, 
and the data by standard deviations, calculated on the non-modulated, filtered maps, without the outer-most multipoles: 
i.e. calculated within range $[\ell_1+1,\ell_2-1]$. This improves the statistical consistency by several percent to the 
case when the calibration is done within the full considered multipole range $[\ell_1,\ell_2]$. 
This procedure has been actually used for the results given in Fig.~\ref{fig:modulation_power}.

The bottom panel of the Fig.~\ref{fig:modulation_power} suggests that the fiducial power spectrum model used by \cite{2007ApJ...660L..81E}:
$C_\ell^{\rm modulated}/C_\ell^{\rm fid} = a(l/l_0)^b$ with additional freedom allowed by the tilt parameter in the reference power spectrum could result in
smaller residuals, however in principle, a possibly large running could also be needed in order to  account for the
residual, systematical discrepancies.

We also check that the same kind of relation (as depicted in Fig.~\ref{fig:modulation_power}) 
with very similar amplitude of the systematical effects is obtained 
for the case of data de-modulation  (see. section~\ref{sec:method}).

We will test the accuracy of the method to reconstruct the modulation parameters in section ~\ref{sec:tests}.

\subsection{Method}
\label{sec:method}

Our method is based on measurements of variances in two opposing hemispheres, in a sample of 3\,000 GRF V5 and ILC5 
simulations and in the corresponding data. 
We use 96 different directions in the northern hemisphere, that define the axial-symmetry axis of the two hemispherical regions. 
The directions are
defined by the pixel centres of the HEALPIX pixelization scheme with the resolution parameter $n_s=4$.
In case of the analysis with V5 data the hemispherical regions are defined outside the KQ75 sky mask,  
and therefore the number of available pixels in these regions ($N_k$) may vary from one orientation to another.

In each region $k=[1,N_r]$ (here $N_r=2$) we measure: 
\begin{equation}
r_k = \frac{\sigma_k^2}{\sigma^2}
\label{eq:measurement}
\end{equation}
where $\sigma_k^2$ is the variance of the CMB within $k$th region and $\sigma^2$ is the variance of the whole map 
(outside the KQ75 sky mask in case of the V5 data). For the set of $N_r$ regional measurements we define the corresponding 
$\chisq$ value as:
\begin{equation}
\chisq = \sum_{k=1}^{N_r} (r_k-\langle r_k^{\rm sim}\rangle )^2/Var(r_k^{\rm sim})
\label{eq:chisq}
\end{equation}
Note that given that we rely on the local variance measurements it is justified, to the limits to which CMB represents a GRF realization,
and to the extent to which a cosmic covariance is unimportant, that we neglect the off-diagonal terms in the covariance matrix, and
treat the regional variance measurements as independent variates.
Although we realize that the $r_k$ quantities defined in Eq.~\ref{eq:measurement} 
should in principle follow a Fischer F-distribution, it is not clear which distribution a sum given in Eq.~\ref{eq:chisq}
should follow, and as such we reconstruct the likelihood function using a generic $\chisq$ distribution
\footnote{We will discuss and take into account possible consequences of this approximation in section~\ref{sec:significance}.}.\\

We seek for the best fit between the data and GRF simulations in terms of the hemispherical variance distributions $(r_1,r_2)$ by
de-modulating the data under assumptions given in Sect.~\ref{sec:assumptions}.
This approach is therefore a non-standard one, due to the fact that generically it's the simulations that are being fit to the data, 
rather than the data to simulations. However within the approximations given in Sect.~\ref{sec:assumptions},
it is possible to reverse the process, by demodulating data, while  retaining a formal correctness and   
allowing thereby to avoid a time-consuming processing of large number of simulations for each cell of the parameter space.
Under no-noise and no-foregrounds assumptions we rewrite the the Eq.~\ref{eq:observation} as:
\begin{equation}
\begin{array}{lll}
T(\mathbf{\hat n}) &=& \mathbf{B}(\mathbf{\hat n},\mathbf{\hat n'}) \star \Bigl( T_{CMB}(\mathbf{\hat n'}) \bigl(1+ M(\mathbf{\hat n'})\bigr) \Bigr) w(\mathbf{\hat n})
\end{array}
\label{eq:observation2}
\end{equation}
It is clear that apart from the beam smoothing effects, the de-modulation of the observed map $T(\mathbf{\hat n})$ is simply a division by the 
factor $\bigl(1+ M(\mathbf{\hat n'})\bigr)$.
In order to account for the beams, 
using few spherical harmonics transformations (SHT), we pre-process the V5 data and simulations as follows:
\begin{enumerate}
\item downgrade the simulations/data to resolution $n_s=128$
\item SHT analysis of the full sky V5 maps to obtain $\alms{V5}$ coefficients, and deconvolve them with the average V channel beam transfer function
\item SHT synthesis using $\alms{V5,nobeam}$ coefficients to obtain a map including the first 128 multipoles (in order to account for the power leakage from cut sky in point~\ref{item:mask})
\item apply KQ75 sky mask to remove the foregrounds in the deconvolved map (there are no foregrounds in the simulations) \label{item:mask}
\item SHT analysis of the cut-sky deconvolved map to obtain $\alms{V5,nobeam,cutsky}$ coefficients up to $\lmax = 80$
\end{enumerate}
We store the final set of  $\alms{V5,nobeam,cutsky}$ coefficients to produce the beam free maps for any requested range of multipoles.
Note that the  two beams of the V1 and V2 WMAP channels are practically identical, however we still average between them to deconvolve the ILC maps.
In the limit of the highest, considered multipoles the operation of deconvolution has an impact of few percent, as 
compared with the convolved power spectrum, as can be inferred from the shape of the beam transfer function.
Also note, that since we operate in the signal dominated regime there is no danger to artificially blow up the 
high-$\ell(\approx 80)$ multipoles.

De-smoothing by the V band beam transfer function, leads to a power increase at $\ell=80$ by about $10\%$.
This can be easily estimated from the transfer function itself, since the transmittance for $\ell=80$ is about $95\%$,
which in the deconvolved power spectrum translates onto an increase by a factor of $1/0.95^2 =\sim 1.1$

We recall that the noise in the case of V5 is, of course, present in the simulations. 
The approximation of ``no noise'' only means that
we assume that the estimates of the modulation parameter, that modulates the pure CMB component, 
is not much altered by the fact 
that we are actually deconvolving noisy observations, rather than a pure CMB component, 
which in general does not make sense unless the signal strongly dominates the noise. 

In case of the ILC5 data we create an effective ``beam'' transfer function by dividing the fit to the ILC5 power spectra (see.~\ref{sec:data_and_sims} for details)
by the fiducial best fit  $\Lambda CDM$ model power spectrum (\cite{2008arXiv0803.0586D} generated using the mean likelihood parameters): i.e. 
$b_\ell^{\rm eff} = \sqrt{\ClILCfit/\Clfid}$. We divide the ILC5 $\alm$ coefficients ($\ell\leq 80$) by this function to match the ILC5 map power spectrum to 
the pure CMB component power spectrum. We will analyze this data with the 3000 GRF signal-only, full sky CMB simulations, generated with the best fit fiducial power 
spectrum $\Clfid$\footnote{Actually for the data preparation process we only generate the GRF $\alm$ 
coefficients and then generate maps for any requested range of multipoles}.\\

As a result, such preprocessed data sets and simulations (apart from the cut sky effects 
which are identical in the two cases) are consistent with the 
fiducial, best fit, theoretical power spectrum, which we verify experimentally. We can rewrite the Eq.~\ref{eq:observation2} as
\begin{equation}
\begin{array}{lll}
T_{\rm nobeam}(\mathbf{\hat n}) &=&  T_{CMB}(\mathbf{\hat n}) \bigl(1+ M(\mathbf{\hat n})\bigr) w(\mathbf{\hat n})
\end{array}
\label{eq:observation3}
\end{equation}
On the basis of our preprocessed data set we assured that the inferred modulation parameter will correspond to the 
modulation of the pure CMB component
(to the extent where the assumptions given in section~\ref{sec:assumptions}  are valid).\\

We seek for the best-fit modulation map $(1+ M(\mathbf{\hat n}))$ such that if the observations ($T_{\rm nobeam}(\mathbf{\hat n})$) are  divided by it,
the resulting map will yield the best consistency with the GRF simulations in terms of the statistics given in Eq.~\ref{eq:chisq}.

\subsection{Parameter space}
\label{sec:parameter_space}

As was mentioned in the previous section, we use 96 different directions in the northern sky, that define
a set of  orientations of the $\Nr$ regions ($\Nr =2$ for hemispherical regions). The set of regions 
uniformly covers the whole sphere.
The 96 directions define our search space, and the corresponding search parameter that we will call ${\mathbf{\hat n_s}}=\{1..96\}$.

Additionally, we use 192 directions over the full sky, that define the orientation of the de-modulation axis $\mathbf{\hat m}=\{1..192\}$. The directions are
defined by the pixel centres of the HEALPIX pixelization scheme of resolution parameter $n_s=4$. Those that are localized in the northern hemisphere
overlap with the directions defining the regions orientations.
These directions define our modulation orientation space.

In the most general case we probe the likelihood function for the modulation amplitudes 
in range $\Alminlmax\in[0.0, 0.2]$ with step $\Delta=0.01$,
and in range $\Alminlmax\in[0.2, 0.3]$ with step $\Delta=0.02$, 
and in range $\Alminlmax\in[0.3, 0.5]$ with step $\Delta=0.05$, 
and in range $\Alminlmax\in[0.5, 0.7]$ with step $\Delta=0.1$.
These values define our modulation amplitude space.

As will be shown in Sect.~\ref{sec:results} including the large modulations ($A\gtrsim 0.5$)
mostly explores completely unimportant regions of the likelihood function, which is why our grid in this region is much sparser. 
In general however, the
amount of the possible hemispherical variance asymmetry in the GRF simulations depends on both:
the underlying power spectrum shape, and the selected range of multipoles.

Additionally we perform search in different bins of multipoles $(\lmin,\lmax)$.
The range of the multipoles tested is summarized in Table~\ref{tab:multipole_range}.
\begin{table}[!h]
\caption{Summary of the tested multipole bins. Note that as explained in section~\ref{sec:modulated_power_spectrum}, 
in the actual analysis we discard the outermost multipoles of the considered multipole ranges.
For clarity, in each cell, we explicitly write down the filtered multipole ranges used. 
The numbers in round brackets indicate the percentage of the variance within a considered range out of the total power 
in the best fit $\Lambda$CDM model, calculated as explained in the text.
The numbers given in square brackets indicate the percentage of variance within a considered range of multipoles 
out of the total CMB signal variance within the first 80 multipoles of the fiducial power spectrum $\Clfid$.}
\centering
\begin{tabular}{c|cccccc}\hline\hline
$\lmin \backslash \lmax$ & 7 & 20 & 30 & 40 & 60 & 80\\
2  & $3\sim 6$ (6.8) [21.2] & $3\sim 19$ (13.8) [43.1] & $3\sim 29$  (16.9) [52.6] & $3\sim 39$  (19.4) [60.4] & $3\sim 59$  (23.8) [74.2] & $3\sim 79$  (27.9) [87.2] \\
6  &                        & $7\sim 19$  (7.0) [21.9] & $7\sim 29$  (10.1) [31.4] & $7\sim 39$  (12.6) [39.2] & $7\sim 59$  (17.0) [53.0] & $7\sim 79$  (21.1) [66.0] \\
19 &                        &                          & $20\sim 29$  (3.0) [9.5]  & $20\sim 39$  (5.6) [17.3] & $20\sim 59$ (10.0) [31.1] & $20\sim 79$ (14.1) [44.1] \\
29 &                        &                          &                           & $30\sim 39$  (2.5) [7.8]  & $30\sim 59$  (6.9) [21.6] & $30\sim 79$ (11.1) [34.6] \\
39 &                        &                          &                           &                           & $40\sim 59$  (4.4) [13.8] & $40\sim 79$  (8.6) [26.8] \\
59 &                        &                          &                           &                           &                           & $60\sim 79$  (4.2) [13.0]
\end{tabular}
\label{tab:multipole_range}
\end{table}

\par According to the CMB WMAP5 best fit $\Lambda$CDM model
\footnote{\href{http://lambda.gsfc.nasa.gov/data/map/dr3/dcp/params/c\_l/wmap\_lcdm\_sz\_lens\_wmap5\_cl\_v3.dat}{http://lambda.gsfc.nasa.gov/data/map/dr3/dcp/params/c\_l/wmap\_lcdm\_sz\_lens\_wmap5\_cl\_v3.dat}}
our considered range of the multipoles: i.e. $\ell \leq 80$ make up for only about 32\% of the total power in this model 
(of which cumulative variance we calibrate to unity at the maximal computed multipole number of $\ell = 2000$:
$\sigma(\lmin,\lmax) = \sum_{\ell=\lmin=2}^{\lmax=80} (2\ell+1) \Clfid / \sum_{\ell=2}^{2000} (2\ell+1) \Clfid$).
However it was shown in \cite{Lew2008} that the modulation of $A\approx 0.1$ extending all the way up to $\lmax =1024$ 
(at which about $96\%$ of the total CMB power is used)
is excluded at a high confidence level ($>99\%$ CL). 

Throughout the analysis we work on maps of the HEALPIX resolution $n_s$, which depends on the considered range of 
multipoles so as to yield the condition: $n_s \geq \lmax/2$.

\subsection{Parameter estimation}
\label{sec:parameter_estimation}
For each direction from our search space (see section~\ref{sec:parameter_space})  we reconstruct the likelihood 
function for each of the the modulation parameters values ${\mathbf \theta}=(A,\mathbf{\hat m})$ and for each considered 
multipole range.
As a first step we a perform minimization of the likelihood over the search parameter ${\mathbf{\hat n_s}}$, in order to select
only the measurements that maximize the possible variance distribution anomaly.
We next derive the corresponding marginal posterior distributions using
flat prior probabilities $\Pi(\mathbf{\theta}|\mathcal{M}) = const.$ 
at each cell of our parameter space. Therefore, the maximum likelihood inference will lead to the same results as the maximal posterior results since according to the Bayes theorem:
\begin{equation}
\mathcal{P}({\mathbf \theta}| \mathcal{M},T(\mathbf{\hat n})) \varpropto \mathcal{L}(T(\mathbf{\hat n})| \mathcal{M},\mathbf{\theta}) \Pi(\mathbf{\theta}|\mathcal{M})
\label{eq:bayes}
\end{equation}
where $\mathcal{P}({\mathbf \theta}| \mathcal{M},T(\mathbf{\hat n}))$ denotes the posterior distribution, and
$\mathcal{L}(\mathbf{\theta}| \mathcal{M},T(\mathbf{\hat n}))$ denotes the likelihood of the parameters $\mathbf{\theta}$
within the hypothesized model $\mathcal{M}$ defined in Eq.~\ref{eq:modulation},.

For the estimates on the modulation amplitude, the marginalized, one-dimensional probability distribution is interpolated using cubic spline interpolation,
before computing the expectancy value, modal value, and confidence ranges. The marginalization over the modulation directions is not performed 
directly on the grid nodes, but rather on an interpolated 
(on the surface of the sphere, for each value of the modulation amplitude independently) posterior.

For the estimates on the modulation orientation, the marginalized, two-dimensional probability distribution is interpolated using two-dimensional
tension splines on sphere \citep{RenkaSSRFPACK}. The marginalization over the modulation amplitude is not performed 
directly on the grid nodes, but rather on an interpolated (for each modulation orientation independently) posterior.
The interpolation is done using a cubic splines and a dense equi-spaced grid.
We also tested and compared the ``interpolations'' using 
spherical harmonics analysis, followed by ${\rm FWHM}=14^\circ$ (the approximate size of the search step) Gaussian smoothing and synthesis up to $\lmax = 30$, and
found that the results: i.e. the maximum likelihood value orientations and confidence contours are reasonably similar, with those where the fitted tension spline 
surface was used.
The usage of cubic interpolations or spherical harmonic approximations in principle can lead to oscillations in the PDF that exceed below the zero value especially in the tails 
of the distribution.
In case of one dimensional interpolations we circumvent this problem by replacing a cubic interpolation with a piece-wise linear interpolation.
Although this step will break the continuity of the first derivative of the PDF function, we mostly probe the likelihood function
dense enough so that these effects are relatively small and unimportant. In case of two-dimensional interpolations we find that the potential oscillations, 
if exist, are small and happen far outside the considered confidence levels. 
In particular such oscillations would lead to artificial generation of multiple isolated contours for a given
CL which we generally do not observe and consequently do not consider this to be a problem.
The most affected artifacts of the applied interpolation are observed for cases where the most preferred 
modulation amplitude is vanishing or is close to zero. In such cases, of course, there is no information on the
modulation orientation.

While deriving our results we choose to rely on dense two-dimensional interpolations using tension splines on sphere 
rather than on,
somewhat arbitrary, spherical harmonics analysis approach. Depending on the tension parameter
the interpolating surface approaches the Delaunay triangulation (linear interpolation) solution for large values of the 
tension parameter, and cubic splines solution for zero-tension parameter. The interpolated surface on an equidistant 
(in galactic latitude and longitude) grid is dense enough so that is could easily be projected without holes 
onto an equal-pixel-area 
HEALPIX grid  to ease the integration over the sphere in pixel space.

\section{Tests of the method}
\label{sec:tests}
To test the correctness of the code, and the sensitivity of our approach we use a GRF white noise simulations. 
The choice of the white noise 
helps to generate a GRF simulation in which the cosmic variance effects, leading to accidental, unequal hemispherical power distribution,
are suppressed, by giving as much power to high-$\ell$ modes, as to the low-$\ell$ modes.
The magnitude of the low-$\ell$ modes dispersion is $\propto \sqrt{N_{\ell}/N_{\ell '}}$; i.e is larger than that for the high-$\ell$ modes, 
where $N_{l\in\{\ell,\ell'\}}=2l+1$ is the number of $\alm$ coefficients at multipoles $\ell,\ell '$ respectively, where  $\ell>\ell '$.
Therefore equalizing power between different multipoles helps to better control the experiment:
i.e. correctly interpret the results of the tests with the synthetic data given some required input modulation parameters.
 
In case of realistic GRF simulations based on the CMB power spectrum, the hemispherical power asymmetry allowed within 
the cosmic variance uncertainty,
is much larger due to the fact that the lowest multipoles (with the smallest number of modes) make up for the main part of the map's total variance, 
while the higher $\ell$ multipoles, even though more numerous, are strongly suppressed.\\

\paragraph{Full sky tests}
In the following tests there are no effects from any instrumental beams, 
nor cut skies. 
We generate a white noise realizations in resolution $n_s=64$ and modulate them with modulation amplitude of $A=0.1$ 
and with modulation axis $\mathbf{\hat m}$ oriented at \lb{225}{-27}. \\

First, we test the correctness of the code by using an analytical proposal for the PDF of the $\chisq$ values.
Since we operate on white noise, zero mean and unit variance GRF simulations,  
its statistical properties are well known, and we therefore approximate the $\chisq$ value 
of the Eq.~\ref{eq:chisq} as:
\begin{equation}
\begin{array}{lll}
\chisq = \sum_{k=1,2} (\sigma^2_k-\langle \sigma^2_k\rangle )^2/Var(\sigma^2_k)
\end{array}
\label{eq:chisq_wn}
\end{equation}
where 
\begin{equation}
\langle \sigma_k^2\rangle  =  \frac{(N_k-1)}{N_k}\sigma^2, \hspace{1cm}  {\rm and} \hspace{1cm} Var(\sigma_k^2)  =  \frac{N^2_k}{2 (N_k-1) }\sigma^4
\label{eq:chisq_th}
\end{equation}
are the expectation value of the mean in the sample, and the 
expected variance of the variances in the sample of variates drawn from the GRF filed \citep{Kenney_variance_stuff}.
The $\sigma^2=1$ is the variance of the Gaussian PDF from which the GRF is drawn, and
$N_k$ is the number of pixels in $k$th region.

\begin{figure*}[!t]
\begin{flushleft}
\begin{tabular}{ll}
\multicolumn{2}{c}{Full sky white-noise maps tests with initial modulation: $A=0.1$ \lb{225}{-27}} \\\\
a) marginalized modulation amplitude PDF: & b) marginalized 50\%, 68\% and 95\% CL modualtion orientation limits\\
\includegraphics[width=0.49\textwidth]{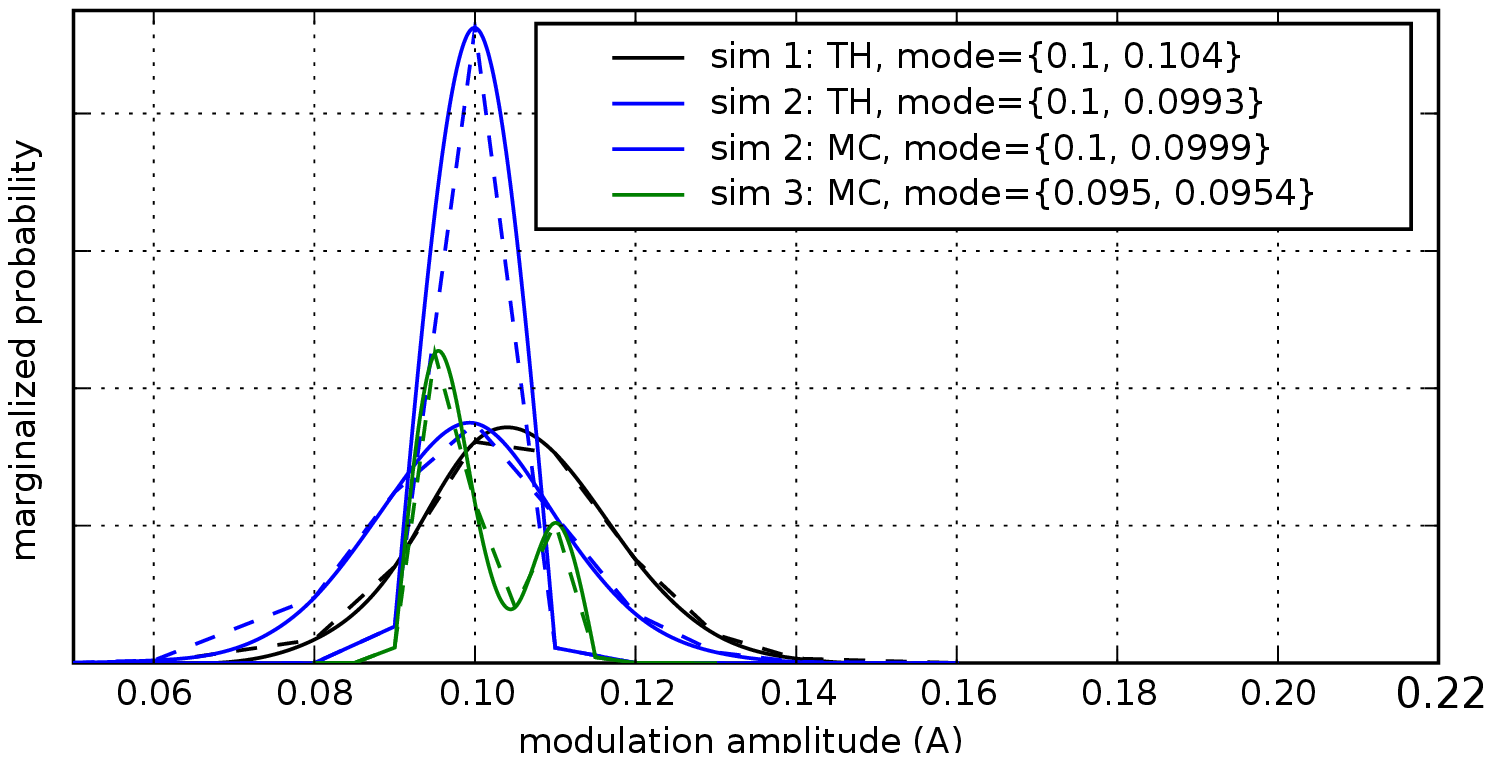} &
\includegraphics[width=0.49\textwidth]{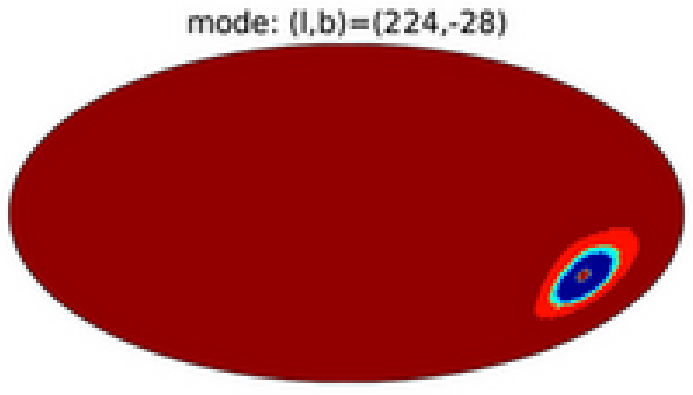}\\\\
\multicolumn{2}{c}{Cut sky (KQ75) white-noise maps tests with initial modulation: $A=0.1$ \lb{225}{-27}} \\\\
c) marginalized modulation amplitude PDF: & d) marginalized 50\%, 68\% and 95\% CL modualtion orientation limits\\
\includegraphics[width=0.49\textwidth]{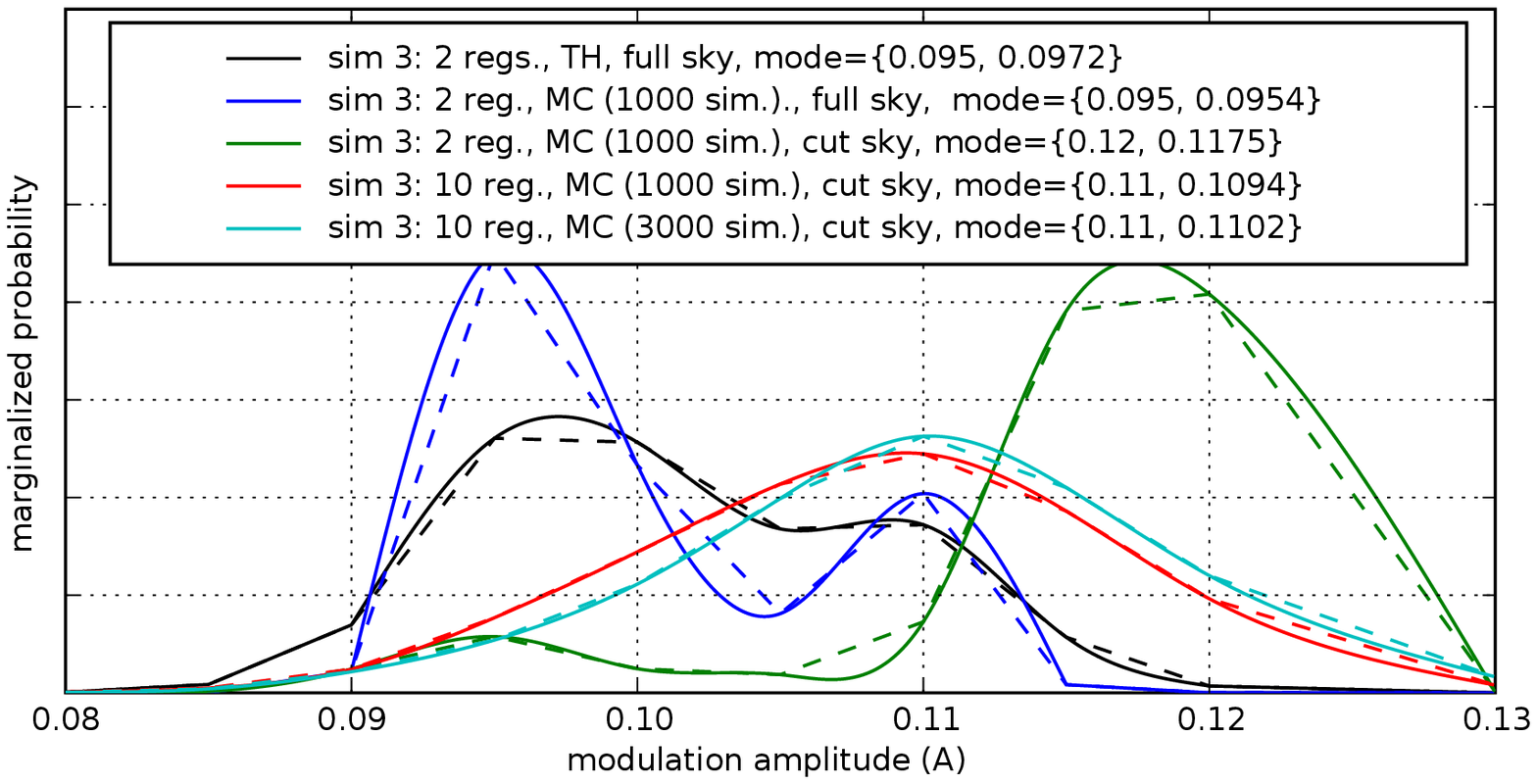} &
\includegraphics[width=0.49\textwidth]{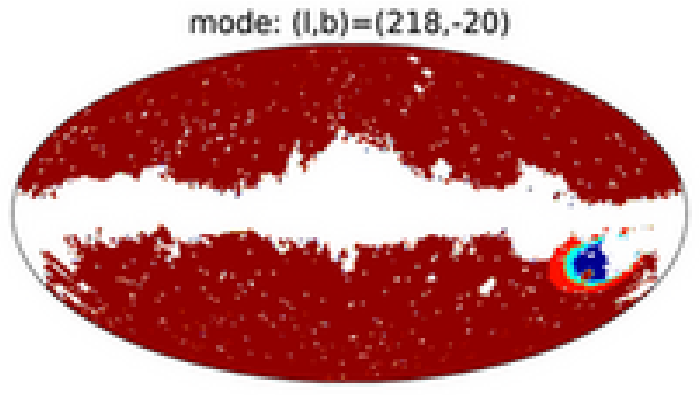}\\\\
\multicolumn{2}{c}{Two-region statistic tests with cut sky (KQ75) white-noise maps and with initial modulation: $A=0.1$ \lb{225}{-27}} \\\\
e) marginalized modulation amplitude PDF: & f) statistics of the reconstructed ML modulation amplitude values \\
\includegraphics[width=0.49\textwidth]{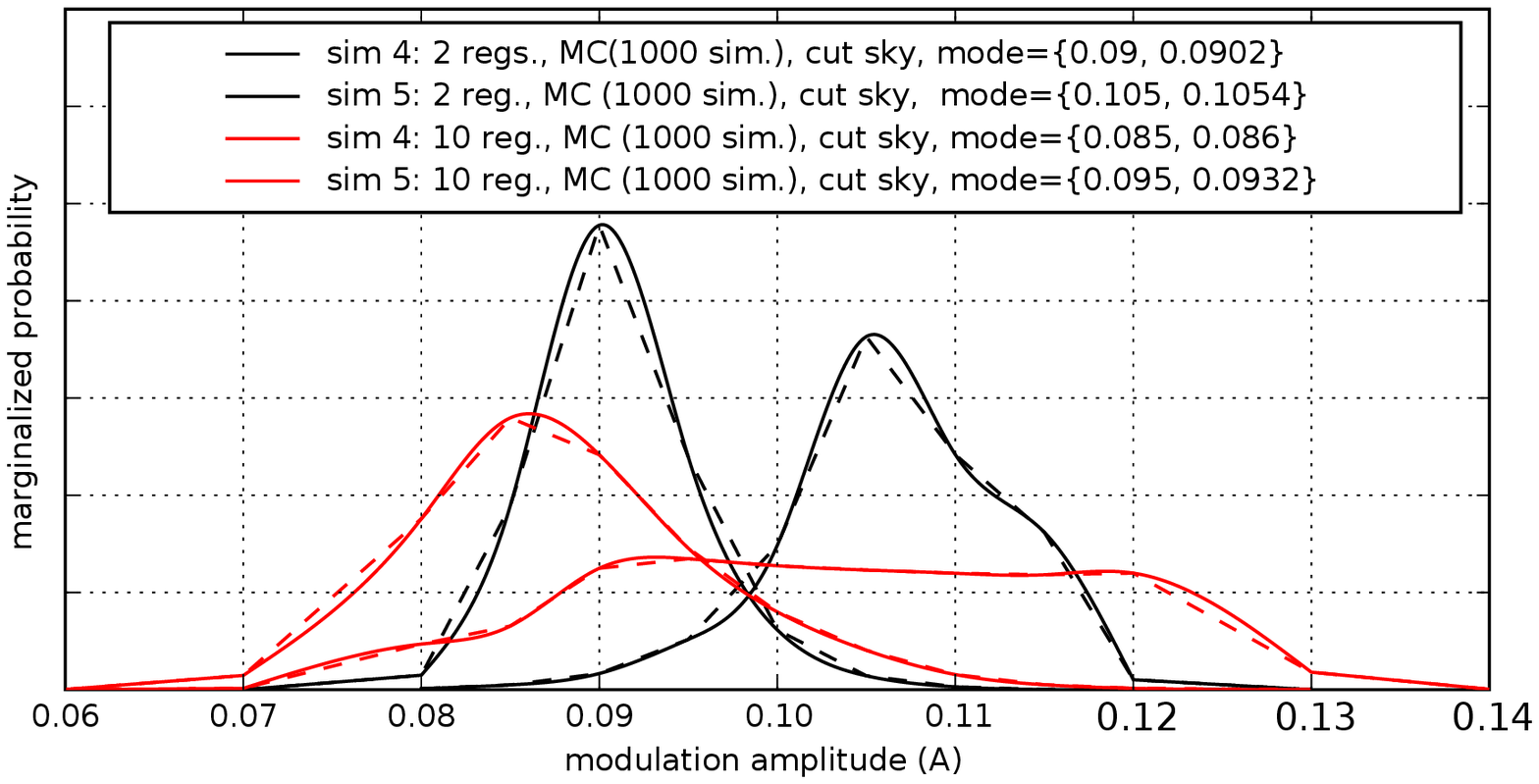} &
\includegraphics[width=0.49\textwidth]{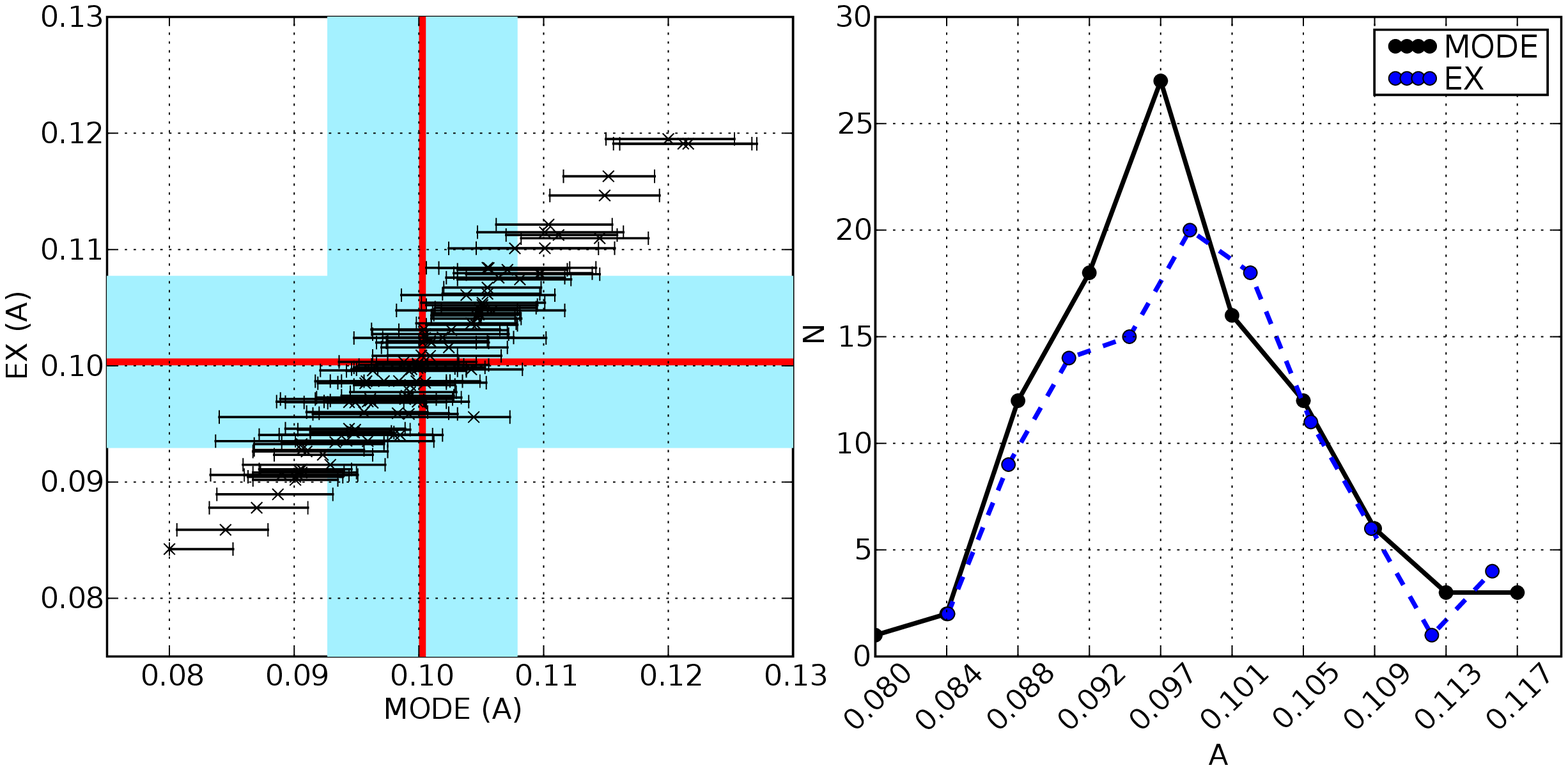}
\end{tabular}
\end{flushleft}
\caption{ White noise maps tests of the accuracy of the method to reconstruct the injected modulation amplitude and its 
orientation. 
{\it (a)} Constraints on the modulation amplitude, from three, chosen, full sky, white-noise, modulated maps. 
The three colors code three selected simulations.
The dashed lines for each simulation connect points at which the likelihood function was calculated, 
and the solid lines represent a cubic polynomial (spline) interpolation, combined with a linear interpolation in the tails of the distribution so as to avoid the oscillations into negative values.
The abbreviations ``TH'' (for the black and the blue-spiky curves) and ``MC'' (for the green and the broad-blue lines) 
in the legend indicate  respectively: the theoretically derived, and Monte-Carlo probed
values of the expected means and variances of the regional (hemispherical) variances (Eq.~\ref{eq:chisq_th}).
In the ``MC'' case a sample of $1\, 000$ simulations is used. The mode values of the probed and interpolated PDFs are also given.
{\it (b)} Limits of the modulation orientation, from one of the modulated, white noise, full sky, GRF realization.
The 50\% (dark blue) , 68\% (light blue) and 95\% (red) confidence level ranges are plotted. 
The small red dot indicates the location of the $\mathbf{\hat m}$ direction, and in the top of the plot
the ML modulation direction is given. We obtain similar results in all tested cases.
{\it (c)} As in panel {\it (a)} but for the KQ75 cut-sky, modulated, white-noise maps tests. 
We focus here only at the peculiar, ``worst-case'' -- the ``simulation 3'', subject to different statistical approaches
as indicated in the legend.
For comparison, the ``TH'' PDF, and the PDF for the full sky case for this simulation, are also replotted.
{\it (d)} Limits of the modulation orientation for the sky cut ``simulation 3'', using $N_r=10$ region statistic. 
The details are as in panel {\it (b)}.
{\it (e)} Constraints on the modulation amplitude, from two another (less-extreme than ``simulation 3'') 
simulations, derived using the two-region statistics (black curves), 
and the corresponding PDFs for 10 regions statistics (red curves). 
{\it (f)} Left: reconstructed modulation PDF mode values versus expectation values from a 100 cut-sky simulations, 
with 68\% CL error bars derived individually from each PDF independently, plotted with the corresponding 
histograms (right). The CL ranges were integrated from the mode value. The red lines indicate the mean 
value of the scatter for each direction, and the shaded area encompasses one standard deviation of the distribution.  
}
\label{fig:tests_wn}
\end{figure*}

We find that the statistics correctly reproduced the initial modulation amplitude and orientation well within the ``one-sigma'' confidence level 
(Fig.~\ref{fig:tests_wn}{\it a}) in all tested cases.
Also, the tests show that using the white noise simulations, the method is able to reject the hypothesis of statistical
isotropy at a very high confidence level (at $\gg 4\sigma$ CL in this case).
The precision of the reconstructed, via interpolation, modulation direction orientation is surprisingly good ($\sim 2^\circ$) given a poor resolution of the search: 
$\sim 14^\circ$ (Fig.~\ref{fig:tests_wn}{\it b})

We also test the reconstruction of the modulation parameters using a set of $1000$ GRF simulations from which we derive the 
averages and variances of the regional variance realizations, and proceed according to equations~\ref{eq:measurement} and~\ref{eq:chisq} for  $N_r=2$. 
We notice an increase in the peakedness of the PDF
when the Monte-Carlo (``MC'') probed estimates are used, as compared to the theoretically (``TH'') derived estimates.
We speculate that the difference might come from the fact that we 
assumed the field variance value $\sigma$ to be unity in all cases.
In practice we will always rely on a sample of $3000$ MC simulations for estimates
of local variance distributions.

Note that the simulation 3 in Fig.~\ref{fig:tests_wn}a, plotted as a peculiarity, found in one of our tests, 
traces the correct value of the injected modulation ($A=0.1$) with an accuracy of about 5\%.

\paragraph{Cut-sky tests and other subtleties}
We further test the stability of the method while varying the number of simulations used to probe the hemispherical (regional)  mean and 
variance expectations. We check the dependence on the increase of the number of simulations from 1000 to 3000.
We perform tests with the KQ75 sky cut, and test sensitivity of the method using different number of 
regions: $N_r=\{2,10\}$. The regions for $N_r>2$ are defined as an
axial-symmetric patches, equally dividing galactic latitude into a symmetrical about the equator (before rotation) regions.
In principle the increased number of regions could potentially have impact on the accuracy of the method.
We use the effective number of degrees of freedom equal to the number of unmasked regions, in order to derive the likelihood value.\\

The results of the tests are presented in Figs.~\ref{fig:tests_wn}{\it c-f}.
We find that the use of the increased number of simulations does not significantly influence the estimates
of the reconstructed mode values, nor the shape of the marginalized PDF.
Also it is clear that the increase of the  number of regions, used in the statistics, broadens the marginalized PDFs,
making thereby the statistic less sensitive.

Furthermore, we see that the accuracy of the method, in case of the cut-sky maps, is generally found at the level of few, 
up to several (in the worst case) percent of the level of the injected modulation ($A=0.1$), 
which is of the same order as the unknown, initial (resulting from a random realization)
modulation\footnote{We refer to the initial unknown, accidentally unequal power distribution in a GRF white 
noise simulation as a ``modulation'' since it's the modulation amplitude that we measure, but of course 
there's no reason to believe that any modulation effect, 
as defined in this paper, exists in the GRF simulations.} 
of our white noise maps.

We note that the selected and presented ``simulation 3'' is one of the worst cases found in our tests, 
and as such, we give more attention to it in variety of tests summarized in Fig.~\ref{fig:tests_wn}{\it c}
where the reconstructed distribution exhibits bi-modality.
In general however, the simulations result in unimodal distributions, 
like those depicted in Fig.~\ref{fig:tests_wn}{\it e}.

We find that generally, in the presence of the cut sky, the modulation orientation is 
correctly reconstructed within 50\% to 68\% CL contours (Fig.~\ref{fig:tests_wn}{\it d}), regardless of the number of regions used in the 
statistics (2 or 10) however in the worst case simulation (as in case of the ``simulation 3'' with the two-region statistics) 
it is found as far as within the 95\% CL contour.

We also checked the difference between different statistical approaches: i.e. 
between maximization\footnote{In case of maximization over the modulation direction orientation, 
we have used the modal values, found in the fitted, two-dimensional
maps of the likelihood function, for each modulation amplitude. We found this method to improve the smoothness 
of the resulting PDF, since our parameter search space is very sparse - only 192 directions over the entire sky.}  
and marginalization over the modulation orientation.
We find that both - the full sky and cut sky tests yield similar, or almost identical results.
We will show that the situation will not be the same in case of the real CMB data or CMB simulations due to the 
effects we mentioned in the beginning of this section.

Finally we tested the statistical biases of the method under the cut sky conditions 
(Fig.~\ref{fig:tests_wn}{\it f (left)} red lines)  
and found that, within the obtained accuracy, no significant statistical bias is noticed.

\paragraph{Summary}
We find that with our chosen search resolution, 
the method traces the correct solution to within a few percent accuracy for the full sky measurements with 
respect to the injected modulation amplitude
value, and from few up to several percent accuracy ($\lesssim 18\%$) for the cut sky case, with about $68\%$
of the estimates yielding an accuracy better than $\sim 7\%$ (Fig.~\ref{fig:tests_wn}f). 
In terms of the absolute errors of the reconstructed value of the modulation amplitude parameter $A$, for the injected amplitude of $A=0.1$
the errors are roughly an order of magnitude smaller: $\sim 0.005$ and $\sim 0.007$ for the full and cut sky cases respectively.
In case of the larger number of regions the sensitivity of the method is worsened (eg. for the case of $N_r=10$) and therefore in the following analysis 
we will only rely on the two-region statistics.

As for the reconstructed modulation direction, we find that mostly the correct direction is reconstructed 
within $\sim 50\%$  CL limits for the two regions statistics, and well within $50\%$ CL limit for 10 regions 
case for the cut sky and full sky cases. 

It is important to note that even with the white-noise simulations the initial, unknown modulation, 
resulting from random, and unequal distribution of 
power in the sky is at level of $A\lesssim 0.005$, which is of the same order of magnitude
as the accuracy which we obtain in the full sky tests. 

As for the reconstructed modulation orientation, mostly the correct direction is found to be within the $50\%$ to $68\%$ CL limits in case of the cut sky 
reconstructions, while the typical angular size of the $50\%$ and $95\%$ CL contours are $\sim 20^\circ$ and $\sim 35^\circ$ respectively (Fig.~\ref{fig:tests_wn}).
The full-sky reconstructions CL contours are slightly smaller.
Note that the  modulation direction is reconstructed via interpolation to within a few degrees accuracy
for the full sky case with the search resolution of about $14^\circ$, which is surprisingly good.
For the cut sky case the accuracy is approximately at the level of several degrees.

\section{Results}
\label{sec:results}

\subsection{Modulation amplitude}
\label{sec:modulation_amplitude}
\begin{figure}[!t]
\vspace{-0.4cm}
\includegraphics[width=\textwidth]{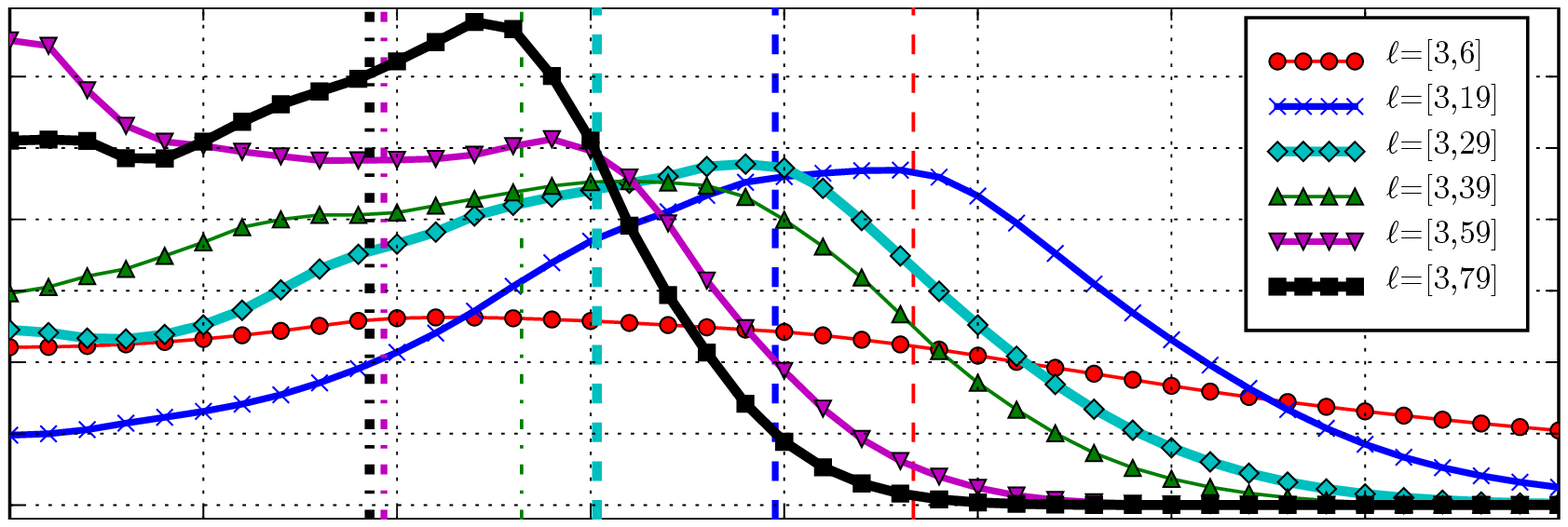}\\
\includegraphics[width=\textwidth]{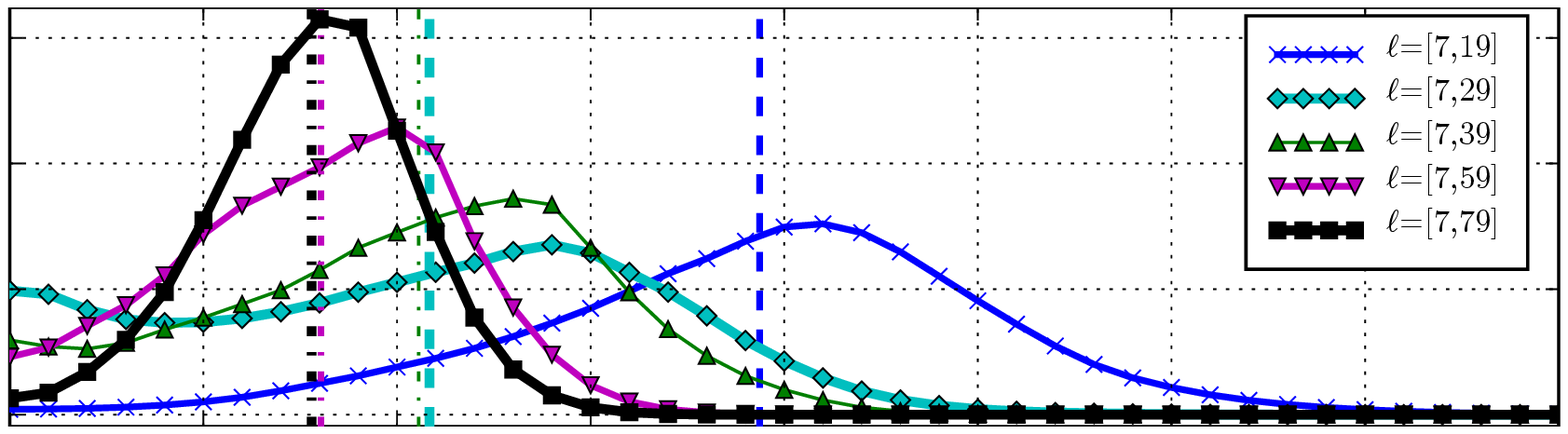}\\
\includegraphics[width=\textwidth]{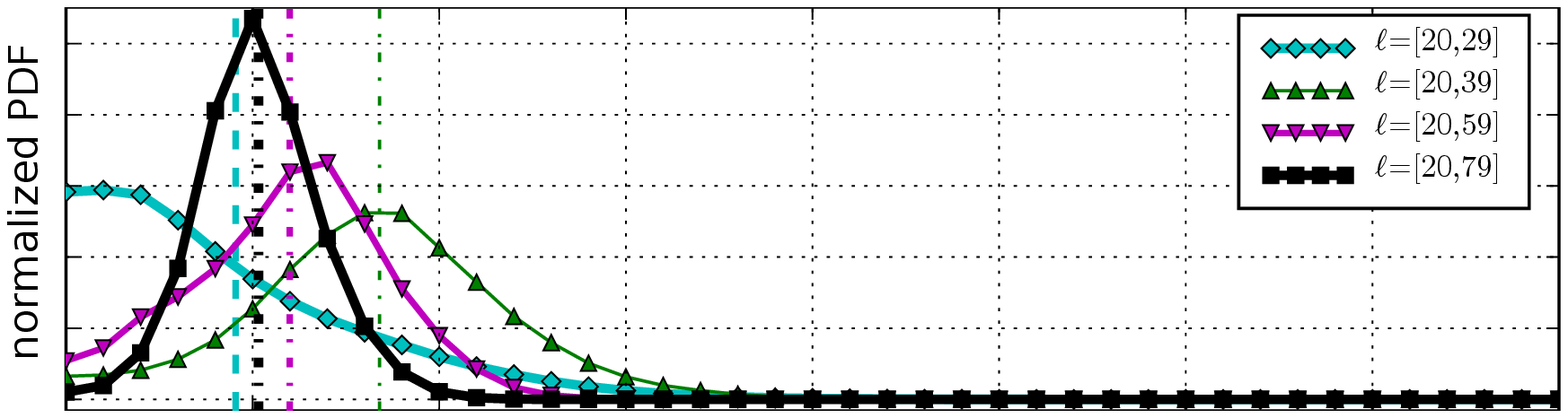}\\
\includegraphics[width=\textwidth]{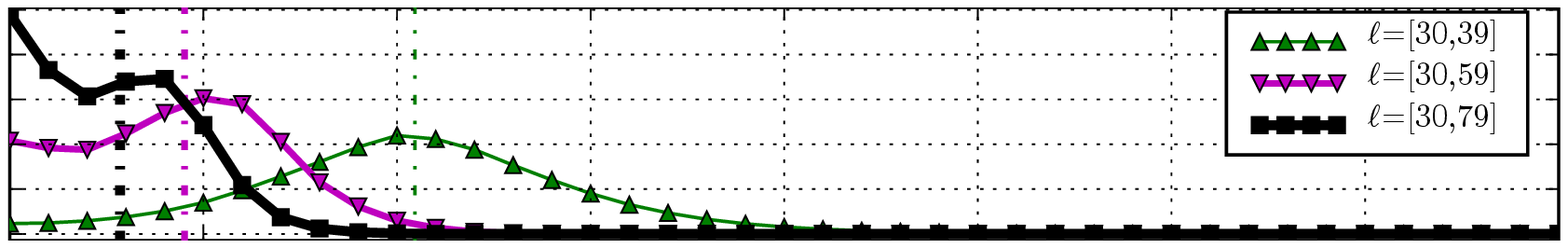}\\
\includegraphics[width=\textwidth]{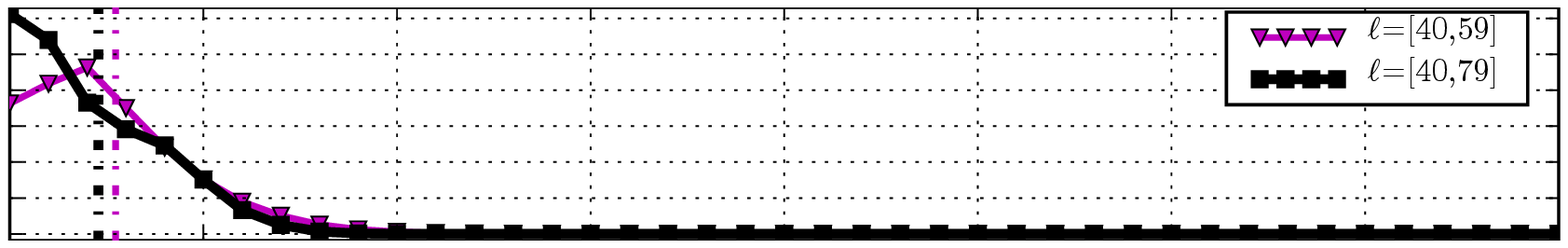}\\
\includegraphics[width=\textwidth]{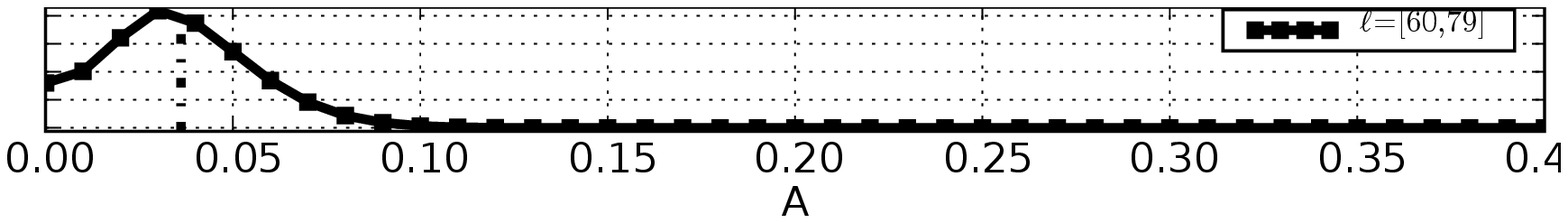}
\caption{Constraints on modulation-amplitude parameters as a function of multipole bin considered, from the V5 data.
From the bottom to the top we plot the data reducing filtering of the large scale multipoles.
For each PDF the corresponding expectancy value is marked by vertical dashed lines for bins $\ell\in[3,6]$, $\ell\in[3,19]$, $\ell\in[3,29]$,
and dash-dotted line for bins $\ell\in[3,39]$, $\ell\in[3,59]$ and $\ell\in[3,79]$. 
Within each group the increasing line width corresponds to increasing value of $\lmax$.
Only every 100th point of the interpolated, marginalized PDF was plotted, so the data points do not correspond to the 
actual grid nodes.
We have truncated the plot at A=0.4 to maximally expose the most interesting regions, 
while keeping the same scale throughout all panels.
}
\label{fig:V5_APDF}
\end{figure}

\begin{figure}[!t]
\includegraphics[width=\textwidth]{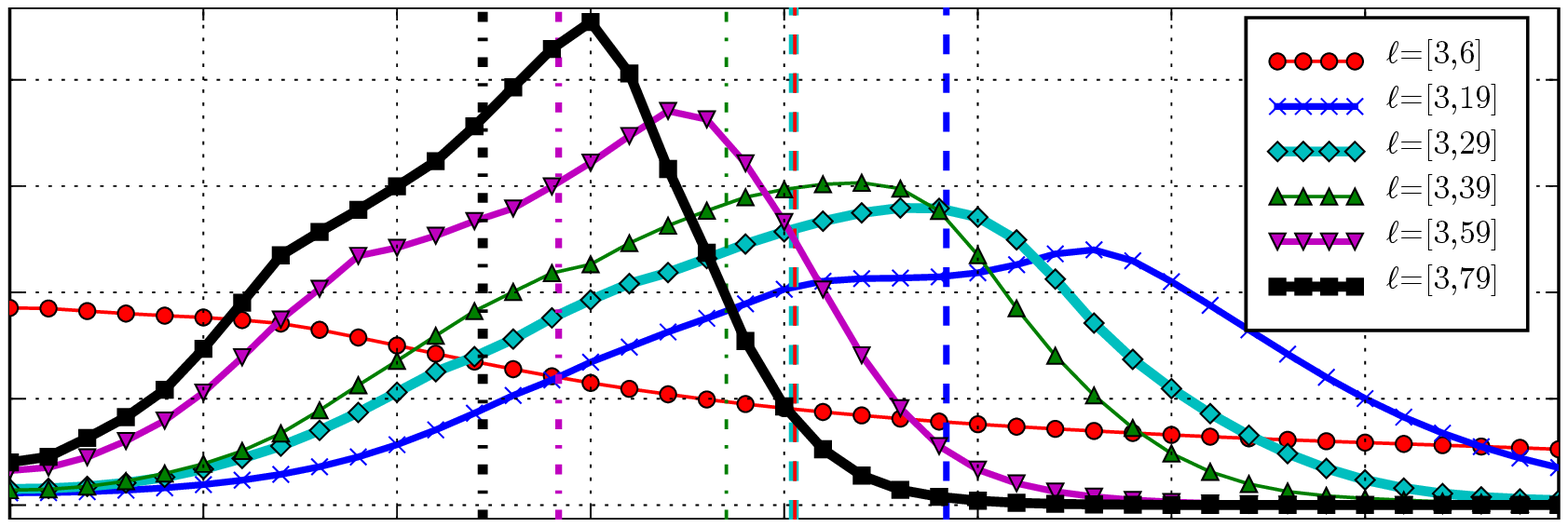}\\
\includegraphics[width=\textwidth]{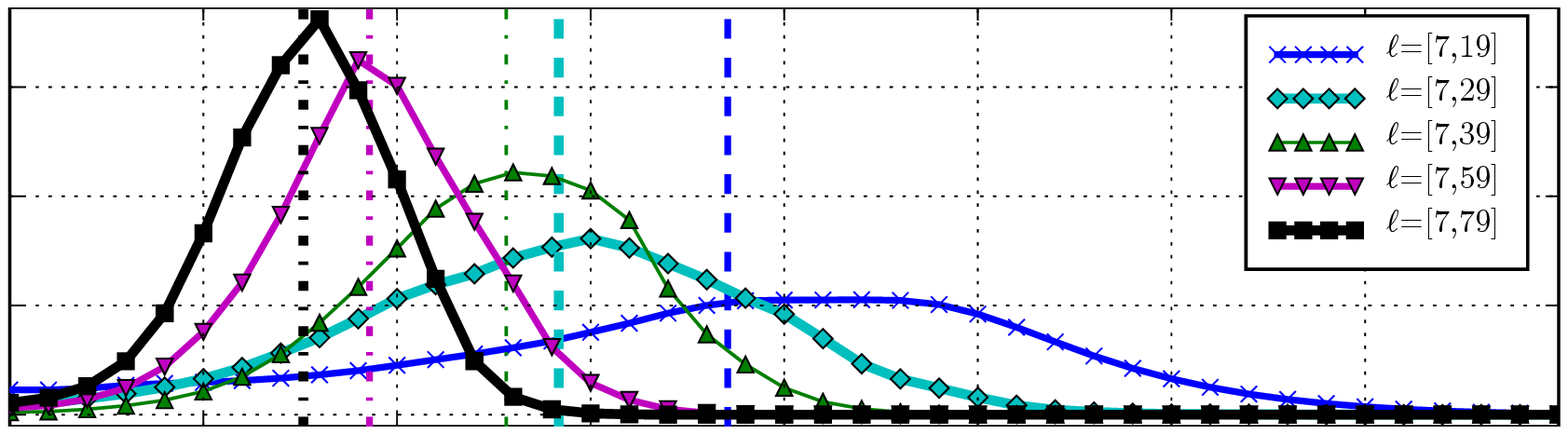}\\
\includegraphics[width=\textwidth]{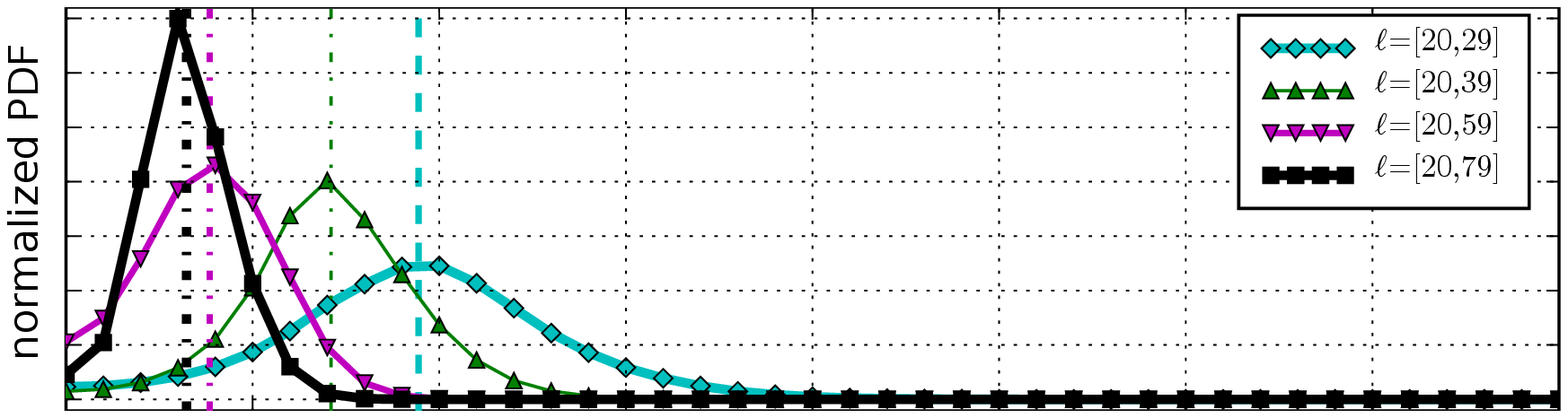}\\
\includegraphics[width=\textwidth]{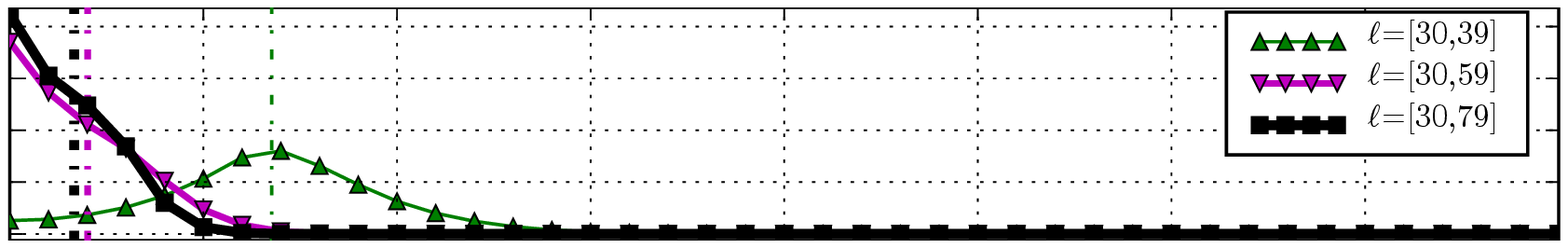}\\
\includegraphics[width=\textwidth]{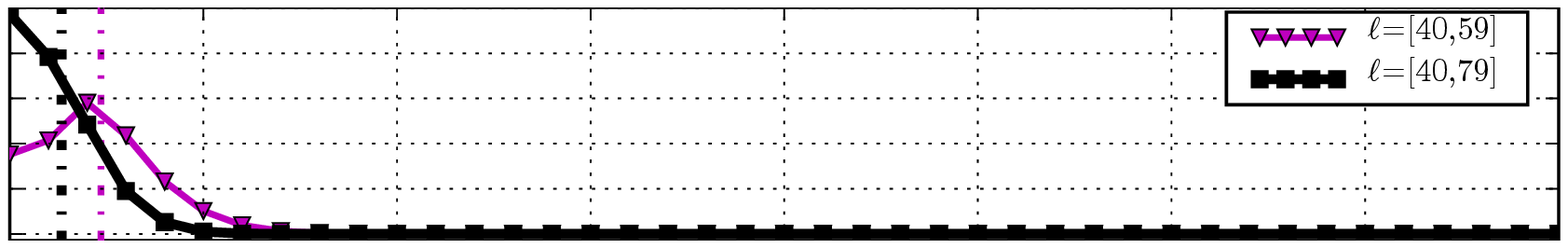}\\
\includegraphics[width=\textwidth]{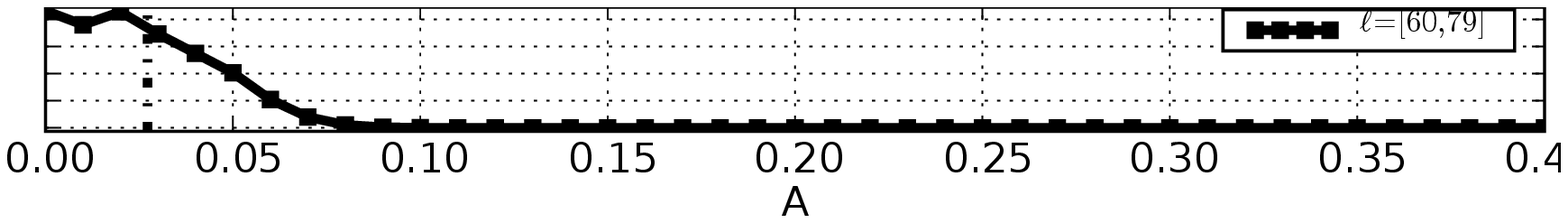}
\caption{Constraints on modulation-amplitude parameter as a function of filtered multipole bin from the ILC5 data.
Details as in Fig.~\ref{fig:V5_APDF}.
}
\label{fig:ILC5_APDF}
\end{figure}

The marginalized over the modulation direction, probability distributions (posteriors) of the 
modulation amplitude parameter as derived from the WMAP V5 and ILC5 data are plotted in 
figures~\ref{fig:V5_APDF} and~\ref{fig:ILC5_APDF} respectively for different ranges of filtered multipoles
(see table~\ref{tab:multipole_range}). 
We remind the reader that only the analysis involving the V5 data was performed using the KQ75 sky mask.

Table~\ref{tab:V5ILC5_ACLr6895} summarizes the modal values of the distributions as parameter estimates and the $68\%$ and $95\%$ confidence level limits.
\begin{table}[!h]
\caption{Constraints on the modulation amplitude parameter for the V5 and ILC5 data.
The table contain the modal parameter values and the corresponding 68\% and 95\% (in brackets) confidence level limits.
}
\centering
\begin{tiny}
\begin{tabular}{c|cccccc}\hline\hline
\multicolumn{7}{c}{V5 data}\\
$\lmin \backslash \lmax$ & 7 & 20 & 30 & 40 & 60 & 80\\
2 & (0.00) 0.00 < 0.11 < 0.29 (0.55) & (0.01) 0.12 < 0.23 < 0.29 (0.34) & (0.00) 0.07 < 0.19 < 0.24 (0.28) & (0.00) 0.05 < 0.16 < 0.21 (0.25) & (0.00) 0.00 < 0.00 < 0.15 (0.20) & (0.00) 0.00 < 0.12 < 0.15 (0.18) \\
6 && (0.07) 0.14 < 0.21 < 0.26 (0.31) & (0.00) 0.00 < 0.14 < 0.18 (0.20) & (0.00) 0.06 < 0.13 < 0.16 (0.18) & (0.01) 0.05 < 0.10 < 0.12 (0.14) & (0.02) 0.05 < 0.08 < 0.11 (0.13) \\
19 &&& (0.00) 0.00 < 0.01 < 0.06 (0.12) & (0.01) 0.05 < 0.09 < 0.12 (0.15) & (0.01) 0.04 < 0.07 < 0.09 (0.11) & (0.02) 0.04 < 0.05 < 0.07 (0.09) \\
29 &&&& (0.01) 0.06 < 0.10 < 0.14 (0.19) & (0.00) 0.00 < 0.05 < 0.07 (0.09) & (0.00) 0.00 < 0.00 < 0.04 (0.06) \\
39 &&&&& (0.00) 0.00 < 0.02 < 0.03 (0.07) & (0.00) 0.00 < 0.00 < 0.03 (0.06) \\
59 &&&&&& (0.00) 0.01 < 0.03 < 0.05 (0.07) \\
\multicolumn{7}{c}{ILC5 data}\\
$\lmin \backslash \lmax$ & 7 & 20 & 30 & 40 & 60 & 80\\
2 & (0.00) 0.00 < 0.00 < 0.26 (0.52) & (0.07) 0.16 < 0.28 < 0.33 (0.41) & (0.06) 0.14 < 0.23 < 0.28 (0.33) & (0.06) 0.13 < 0.22 < 0.26 (0.30) & (0.03) 0.09 < 0.17 < 0.20 (0.24) & (0.03) 0.08 < 0.15 < 0.18 (0.20) \\
6 && (0.01) 0.12 < 0.22 < 0.28 (0.31) & (0.04) 0.09 < 0.15 < 0.20 (0.24) & (0.06) 0.10 < 0.13 < 0.17 (0.20) & (0.04) 0.07 < 0.09 < 0.12 (0.15) & (0.03) 0.05 < 0.08 < 0.10 (0.12) \\
19 &&& (0.02) 0.06 < 0.10 < 0.13 (0.17) & (0.02) 0.05 < 0.07 < 0.09 (0.12) & (0.00) 0.02 < 0.04 < 0.06 (0.07) & (0.01) 0.02 < 0.03 < 0.04 (0.06) \\
29 &&&& (0.00) 0.04 < 0.07 < 0.10 (0.12) & (0.00) 0.00 < 0.00 < 0.03 (0.05) & (0.00) 0.00 < 0.00 < 0.02 (0.04) \\
39 &&&&& (0.00) 0.01 < 0.02 < 0.03 (0.05) & (0.00) 0.00 < 0.00 < 0.02 (0.03) \\
59 &&&&&& (0.00) 0.00 < 0.00 < 0.03 (0.06) \\
\end{tabular}
\end{tiny}
\label{tab:V5ILC5_ACLr6895}
\end{table}

In Table~\ref{tab:V5ILC5_CLrA0} we concisely summarize the results of the modulation significance analysis:
ie. the analysis in which we derive the minimal confidence levels, at which  the modulation value 
of $A=0$ cannot be excluded.
We specify the expectancy values, mode values of the distributions,
and the corresponding significance. 
We choose to calculate the confidence intervals - or rather, since we're working on posterior probability distributions, 
in the nomenclature of the Bayesian language, the credibility intervals, 
by integrating from the modal value, rather than from the expectancy value.

\begin{table}[!h]
\caption{Results of the modulation amplitude parameter estimation for the V5 and ILC5 dataset.
The table contain the minimal confidence levels (as percents) at which the parameter value of $A=0$ cannot be excluded 
(bold face numbers) and the  expectancy (in round brackets) and the modal (in square brackets) values of the 
corresponding distributions. See also Fig.~\ref{fig:V5ILC5_CLrA0}.
}
\centering
\begin{tabular}{c|cccccc}\hline\hline
\multicolumn{7}{c}{V5 data}\\
$\lmin \backslash \lmax$ & 7 & 20 & 30 & 40 & 60 & 80\\
2 & \textbf{57.5}  (0.23) [0.11] & \textbf{96.1}  (0.20) [0.23] & \textbf{90.9}  (0.15) [0.19] & \textbf{88.7}  (0.13) [0.16] & \textbf{4.6}  (0.10) [0.00] & \textbf{86.4}  (0.09) [0.12] \\
6 && \textbf{99.5}  (0.19) [0.21] & \textbf{90.2}  (0.11) [0.14] & \textbf{94.5}  (0.11) [0.13] & \textbf{97.0}  (0.08) [0.10] & \textbf{99.4}  (0.08) [0.08] \\
19 &&& \textbf{25.8}  (0.05) [0.01] & \textbf{96.8}  (0.08) [0.09] & \textbf{97.2}  (0.06) [0.07] & \textbf{99.7}  (0.05) [0.05] \\
29 &&&& \textbf{97.1}  (0.10) [0.10] & \textbf{86.0}  (0.05) [0.05] & \textbf{0.2}  (0.03) [0.00] \\
39 &&&&& \textbf{61.1}  (0.03) [0.02] & \textbf{0.3}  (0.02) [0.00] \\
59 &&&&&& \textbf{88.8}  (0.04) [0.03] \\
\multicolumn{7}{c}{ILC5 data}\\
$\lmin \backslash \lmax$ & 7 & 20 & 30 & 40 & 60 & 80\\
2 & \textbf{0.4}  (0.20) [0.00] & \textbf{99.1}  (0.24) [0.28] & \textbf{99.1}  (0.20) [0.23] & \textbf{99.3}  (0.19) [0.22] & \textbf{98.7}  (0.14) [0.17] & \textbf{98.8}  (0.12) [0.15] \\
6 && \textbf{96.6}  (0.19) [0.22] & \textbf{99.2}  (0.14) [0.15] & \textbf{99.9}  (0.13) [0.13] & \textbf{99.7}  (0.09) [0.09] & \textbf{99.6}  (0.08) [0.08] \\
19 &&& \textbf{97.9}  (0.09) [0.10] & \textbf{99.2}  (0.07) [0.07] & \textbf{95.6}  (0.04) [0.04] & \textbf{98.7}  (0.03) [0.03] \\
29 &&&& \textbf{95.7}  (0.07) [0.07] & \textbf{0.4}  (0.02) [0.00] & \textbf{0.4}  (0.02) [0.00] \\
39 &&&&& \textbf{78.1}  (0.02) [0.02] & \textbf{0.5}  (0.01) [0.00] \\
59 &&&&&& \textbf{0.2}  (0.03) [0.00] \\
\end{tabular}
\label{tab:V5ILC5_CLrA0}
\end{table}

\begin{figure}[!t]
\centering
V5 data\\
\includegraphics[width=\textwidth]{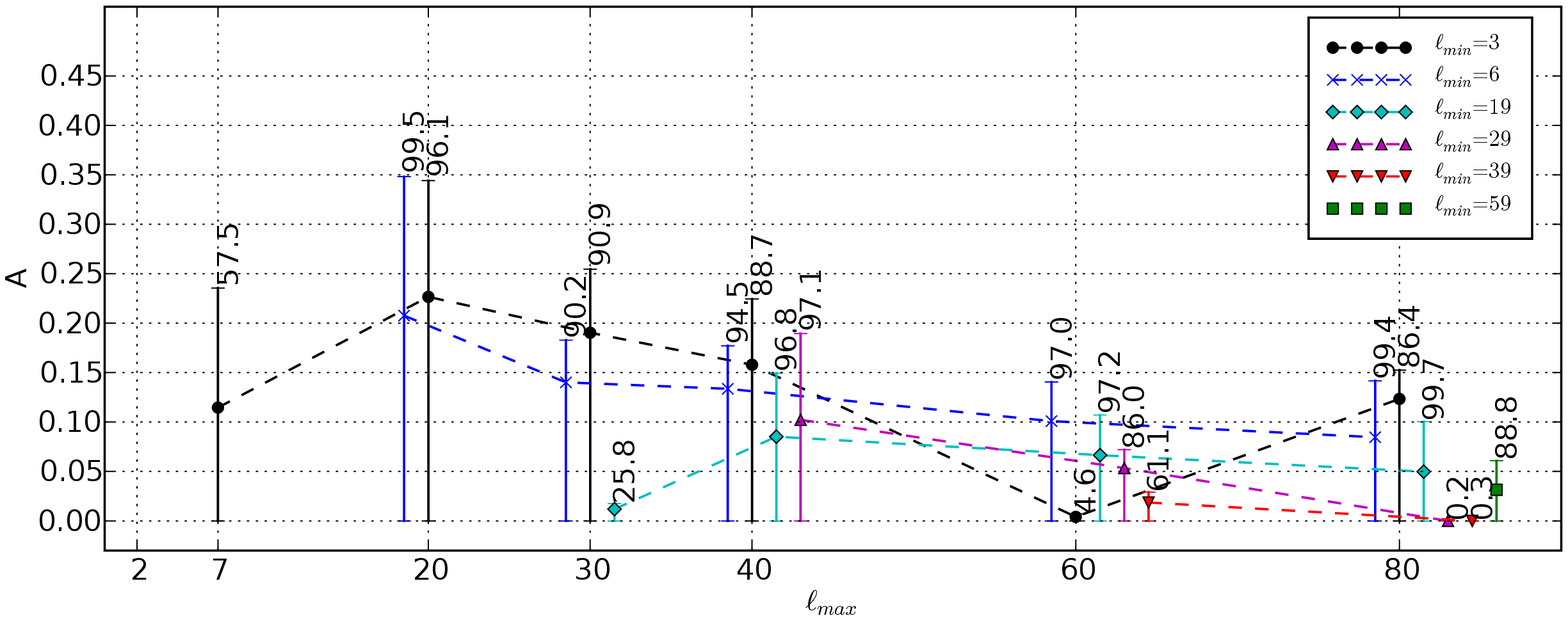}\\
ILC5 data\\
\includegraphics[width=\textwidth]{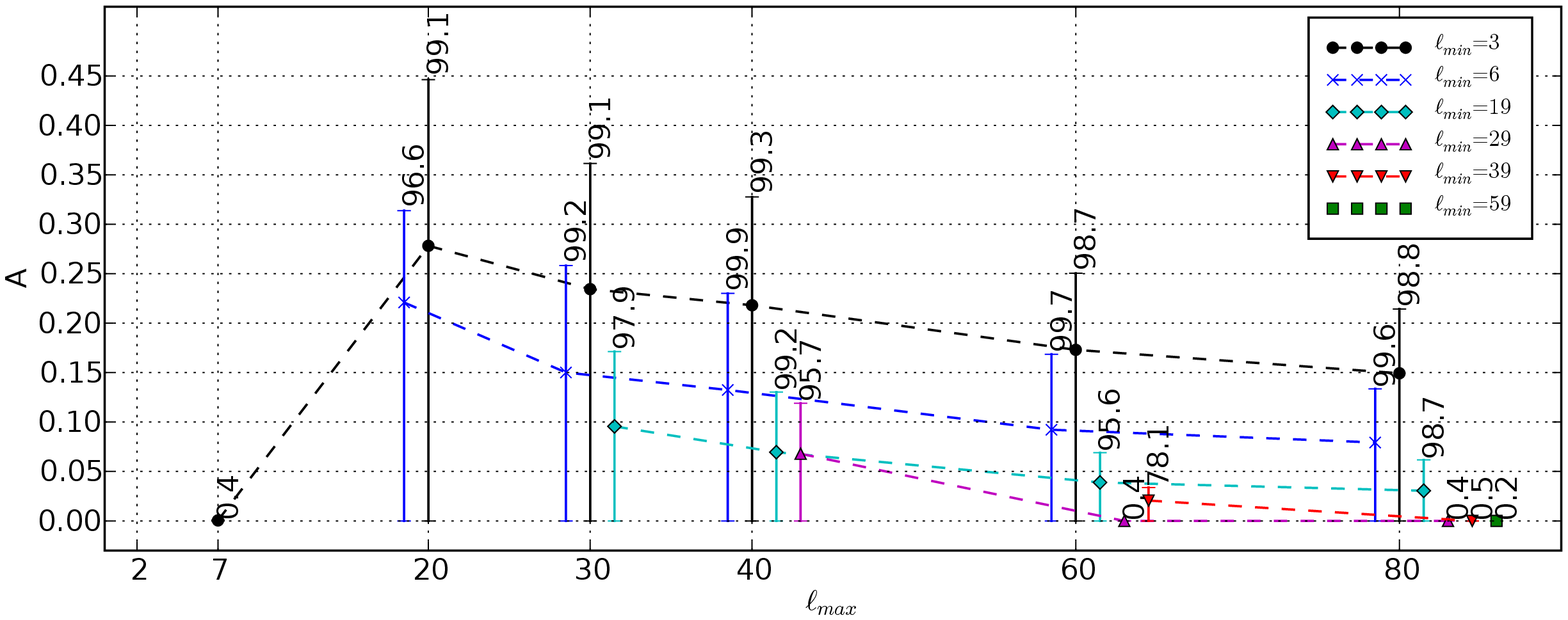}
\caption{Results of the modulation amplitude parameter estimation from the V5 data (top panel) 
and ILC5 data (bottom panel). 
Modal values of the posterior distributions are plotted for different bins of multipoles filtered out from the dataset.
The error bars represent the confidence limits that include the $A=0$ value. The corresponding confidence levels 
(given in percents) are given as annotations by the data points.
}
\label{fig:V5ILC5_CLrA0}
\end{figure}

To visualize these results, we plot the estimated modal values of the posterior distributions as a 
function of considered $\lmin$ and $\lmax$ values (Fig.~\ref{fig:V5ILC5_CLrA0}). 
For each multipole bin, in Fig.~\ref{fig:V5ILC5_CLrA0} we also indicate the minimal confidence 
level (see  Table~\ref{tab:V5ILC5_CLrA0}) it takes to
keep the modulation $A=0$ i.e. the non-modulated, isotropic model, as an  viable option.\\

It is clear that the modulation amplitude depends on the considered multipole range and hence on scale.
Is is generally seen that large values of the best-fit modulations mostly come from the large scales, 
while for high multipole bins, the best-fit modulations are much smaller.
As explained in section~\ref{sec:tests}, within GRF realizations this is somewhat expected 
due to the nature of cosmic variance effects.

Looking at distributions in  Figs.~\ref{fig:V5_APDF} and~\ref{fig:ILC5_APDF},
it is apparent that the modulation for $\ell\in [7,19]$ (and also $\ell\in [3,19]$), in the V5 data,
(blue curves (crosses) in Fig.~\ref{fig:V5_APDF})
is strongest and most significant, as it takes the confidence level as high as $\sim 99.5\%$ to include the $A=0$ value
(see.Table~\ref{tab:V5ILC5_CLrA0} and Fig.~\ref{fig:V5ILC5_CLrA0}).
The appearance of some asymmetry in this range also seems consistent with the results presented in 
Section~\ref{sec:power_ratio}.
The range of multipoles $\ell\in [20,29]$ of the V5 data does not seem to prefer any modulation value as its modal 
value is almost zero, and consequently while increasing values of $\lmax$,
for $\lmin=7$, the overall significance falls, as this multipoles bin is included, 
but then systematically increases as larger multipoles are added,
which is consistent with the shape of the PDF functions preferring some non-zero modulation for higher 
multipole bins like $\ell\in [30,39]$.

The best-fit modulations for $\ell\in [20,39]$ and $\ell\in [30,39]$ 
range from $A=0.07$ to $A=0.10$ and exclude the isotropic model, 
depending on the data, at confidence level of about $96\%$ to $99\%$, while
the best-fit modulation for $\ell\in [7,39]$ in the ILC5 data, with the modal amplitude of $A=0.13$,
exclude the isotropic model, at confidence level as high as $99.9\%$.
We will further test the significance of these results in section~\ref{sec:significance}.

Note also that, some of the marginalized PDFs exhibit bi-modality 
(eg. the PDF corresponding to the multipole range $\ell\in[3,59]$ -- magenta-line ($\triangledown$) in Fig.~\ref{fig:V5_APDF}).
This bi-modality results from the marginalization itself, and is not observed in the full non-marginalized distribution.
Since the likelihood function does not depend on the orientation of the modulation axis for the modulation 
amplitude $A=0$, while it does depend on the modulation orientation very strongly for modulations $A\gg 0$, 
the likelihood surface
shall tend to be peaky for large values of $A$, and flat for $A\approx 0$. Hence, depending on how strong the 
preference of some direction happens to be, it is possible to accumulate in the marginalization process a second 
peak (the second mode) out of somewhat less-preferred, but constant at certain level, likelihood values along  
$A\approx 0$ direction.
As a result the aforementioned range $\ell\in [3,59]$ of V5 data yields a small significance in 
Table~\ref{tab:V5ILC5_CLrA0} (in terms of rejecting an isotropic model).
We have also processed the results by using maximization over the modulation orientations instead of marginalizations, 
and as expected, the maximized PDFs are unimodal and more strongly exclude the non-modulated, isotropic models, 
however we choose the more conservative, and more correct method of marginalizing over the non plotted dimensions.\\

From Fig.~\ref{fig:V5ILC5_CLrA0} it is easy to see that the modulation amplitude estimates are mostly similar 
between the two datasets, and that generally the amplitude of the modulation decreases with increasing multipole 
number $\ell$ in the two datasets.
A common feature between the two datasets is that for high multipole bins $\ell\in [40,59]$, $\ell\in [40,79]$ 
and $\ell\in [60,79]$ the best-fit modulation amplitudes are small or zero, despite that the amount of variance
carried by bin eg.  $\ell\in [40,79]$ is as high as $\sim 27\%$ of the total variance carried
by the full range of considered  multipoles (i.e. from $\ell=3$ to $\ell=79$, see.table~\ref{tab:multipole_range}).
The multipoles range $\ell\in [29,40]$ instead  participate to the total variance only by $\sim 8\%$,  
and consequently,  the the best-fit modulation of this range, estimated to be $A=0.10$ ($A=0.07$)
for the V5 (ILC5) data,
is effectively destroyed, as the higher multipole bins are included,  most likely due to simply
dominating power in the added multipole bins.\\

There are few significant differences between the datasets as well. 
Firstly, we notice, that the ILC5 estimates are generally slightly, but systematically larger from the V5 estimates.
Also, in particular, the full sky ILC5 posteriors, for multipole bins  $\ell\in [7,29]$ and $\ell\in [20,29]$
strongly prefer some non-vanishing modulation amplitude, in contrast to the V5 data 
(compare cyan-diamonds in Fig.~\ref{fig:V5_APDF} and~\ref{fig:ILC5_APDF}). 
This results in an almost constant significance of excluding $A=0$ as higher multipole bins are being included 
(see first three rows of the  Table~\ref{tab:V5ILC5_CLrA0} in section for ILC5 data).

It is interesting to note a small difference in range $\ell\in[3,6]$ (see table~\ref{tab:V5ILC5_CLrA0}) 
in which  the ILC5 slightly  prefer a vanishing best-fit modulation.
Note that among our considered multipole bins, the ILC5 should be reliable basically only within this lowest range. 
In contrast, some non-zero preferred modulation is obtained with the V5 data; however  
the value is still largely consistent with the vanishing modulation at confidence level as low as $\sim 58\%$.\\

Some differences between the datasets are of course expected due to the cut-sky effects, which preclude 
filtering of exactly the same range of multipoles due to the power leakage effects in case of V5 data.
Also caution is needed in the interpretation of the ILC5 data for higher multipole bins, as residual foregrounds
in the regions around the Galactic center, may have some impact on the results.
In particular, these residual foregrounds might be responsible for the significant alteration of the shape of the PDF 
function in the multipole bin $\ell\in [20,29]$  (cyan-diamonds in figures~\ref{fig:V5_APDF} and ~\ref{fig:ILC5_APDF})
towards an increased significance in favor of non-isotropic models.

\subsection{Modulation orientation}
\label{sec:modulation_orientation}
We now focus on the  modulation orientation as a function of our chosen multipole bins, 
as specified in table~\ref{tab:multipole_range}.
The maximum likelihood modulation orientations are summarized in table~\ref{tab:V5ILC5_ML}
\begin{figure}[!htb]
\begin{tabular}{cccccc}
\multicolumn{6}{c}{V5 data}\\
$\lmin=2$ \\
$\lmax=7$ & $\lmax=20$ & $\lmax=30$ & $\lmax=40$ & $\lmax=60$ & $\lmax=80$\\
\includegraphics[width=0.166\textwidth]{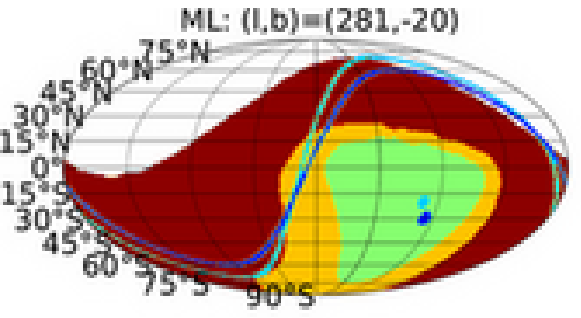}&
\includegraphics[width=0.166\textwidth]{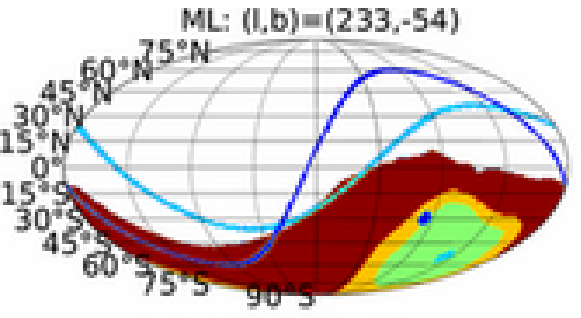}&
\includegraphics[width=0.166\textwidth]{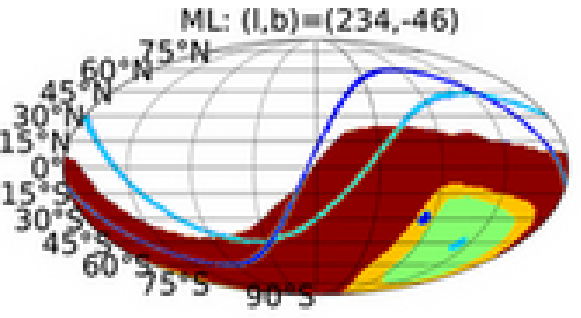}&
\includegraphics[width=0.166\textwidth]{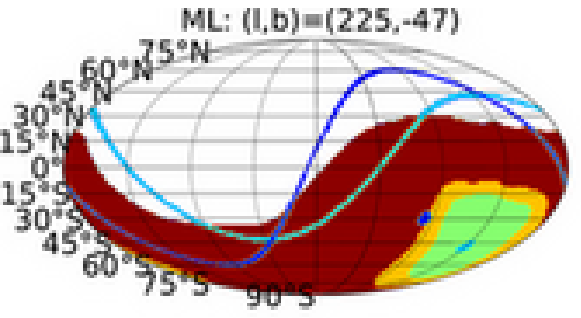}&
\includegraphics[width=0.166\textwidth]{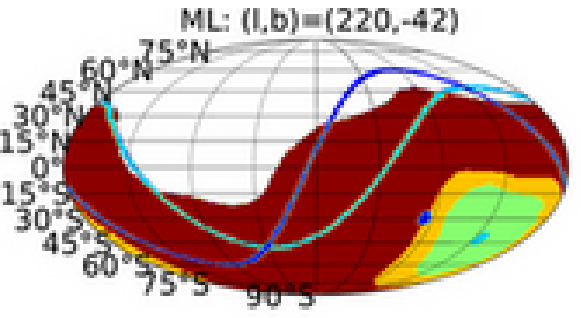}&
\includegraphics[width=0.166\textwidth]{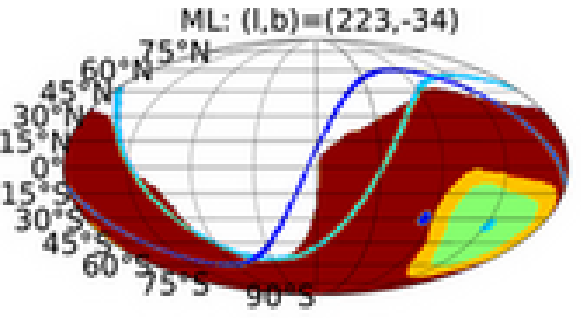}\\
$\lmin=6$ &
\includegraphics[width=0.166\textwidth]{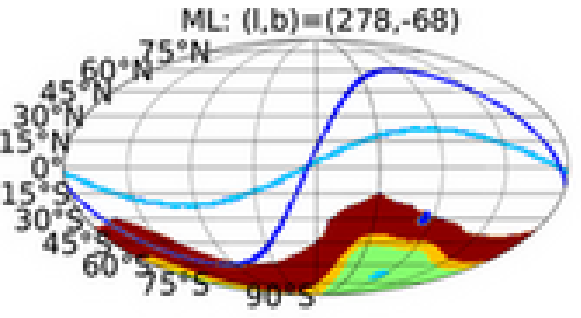}&
\includegraphics[width=0.166\textwidth]{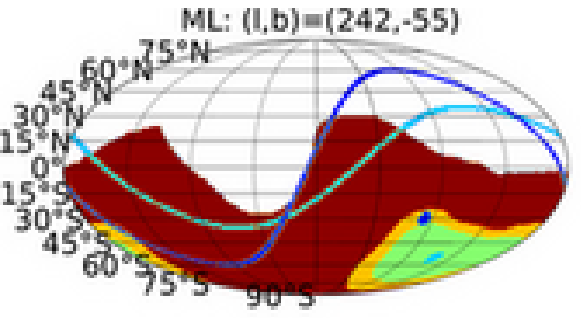}&
\includegraphics[width=0.166\textwidth]{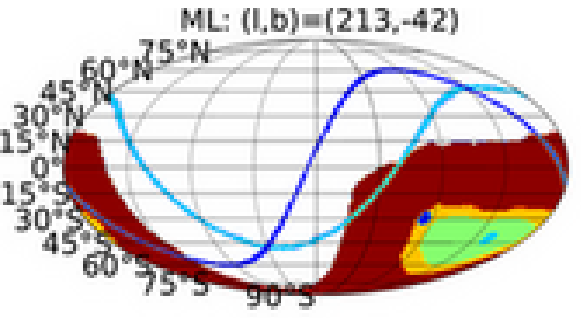}&
\includegraphics[width=0.166\textwidth]{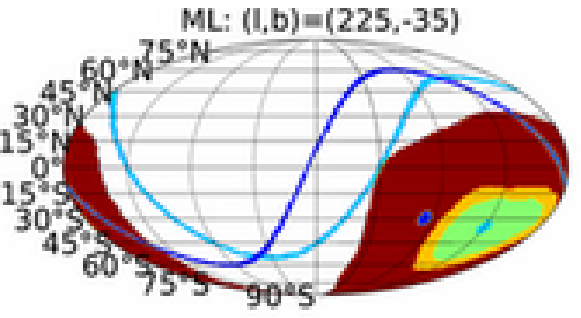}&
\includegraphics[width=0.166\textwidth]{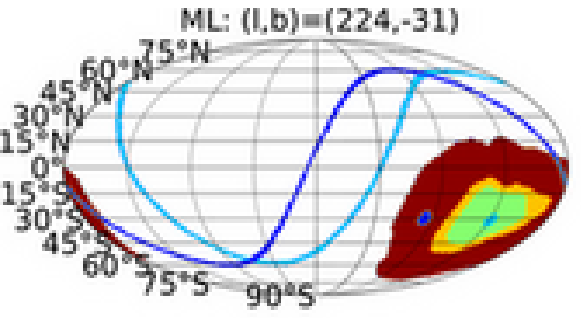}\\
&
$\lmin=19$ &
\includegraphics[width=0.166\textwidth]{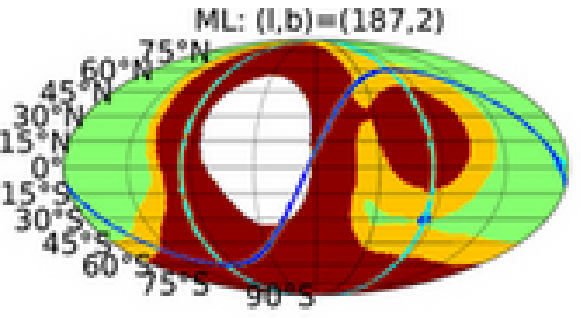}&
\includegraphics[width=0.166\textwidth]{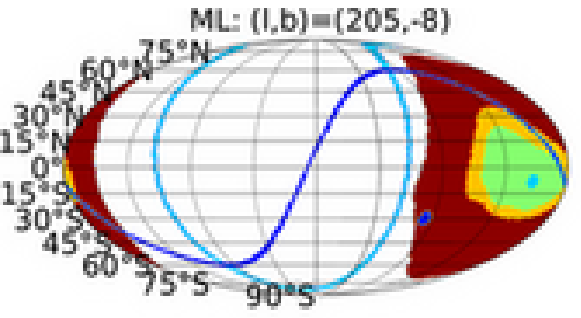}&
\includegraphics[width=0.166\textwidth]{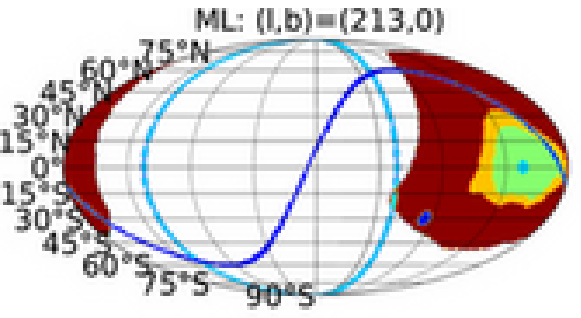}&
\includegraphics[width=0.166\textwidth]{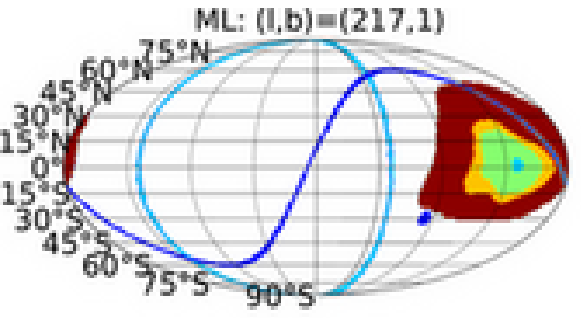}\\
&
&
$\lmin=29$ &
\includegraphics[width=0.166\textwidth]{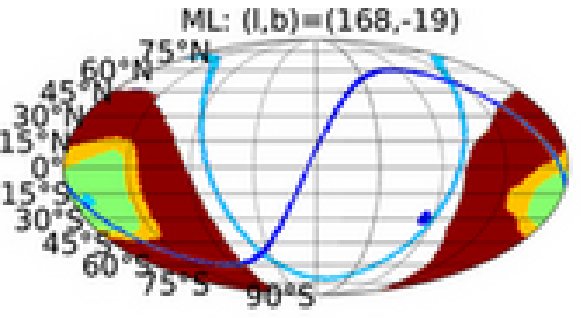}&
\includegraphics[width=0.166\textwidth]{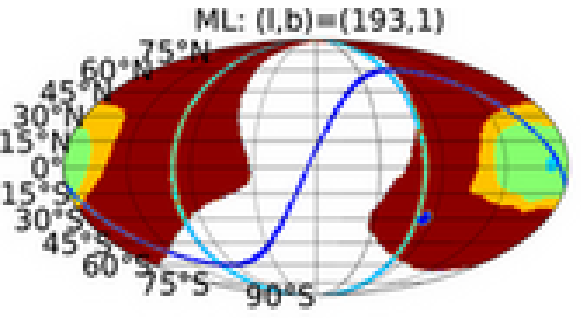}&
\includegraphics[width=0.166\textwidth]{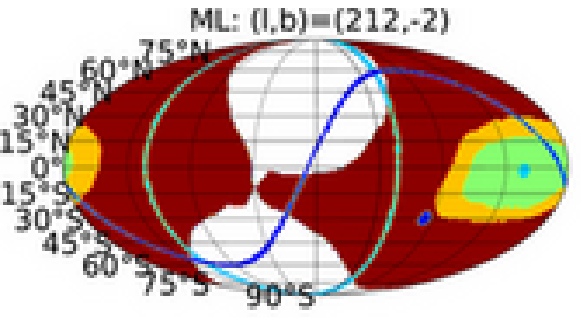}\\
&
&
&
$\lmin=39$ &
\includegraphics[width=0.166\textwidth]{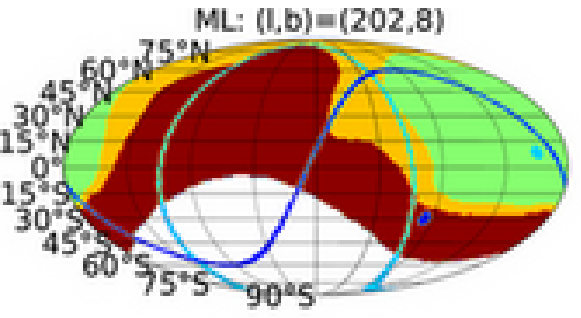}&
\includegraphics[width=0.166\textwidth]{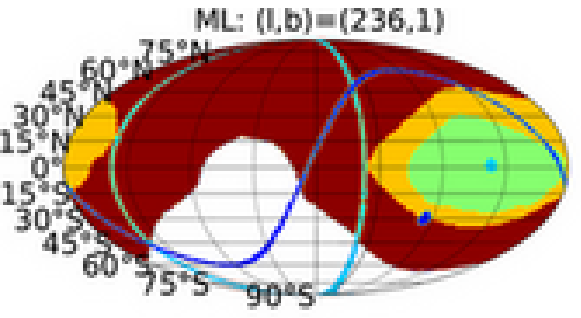}\\
&
&
&
&
$\lmin=59$ &
\includegraphics[width=0.16\textwidth]{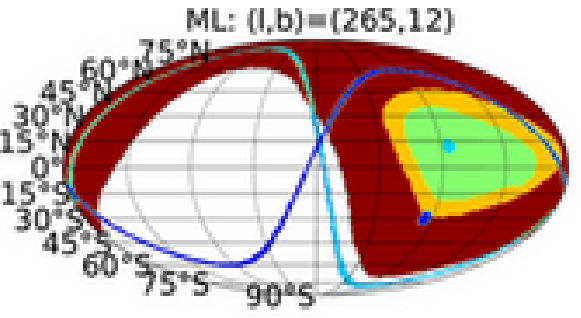}\\
\multicolumn{6}{c}{ILC5 data}\\
$\lmin=2$ \\
$\lmax=7$ & $\lmax=20$ & $\lmax=30$ & $\lmax=40$ & $\lmax=60$ & $\lmax=80$\\
\includegraphics[width=0.166\textwidth]{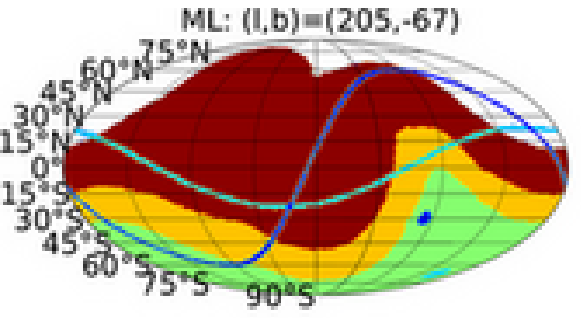}&
\includegraphics[width=0.166\textwidth]{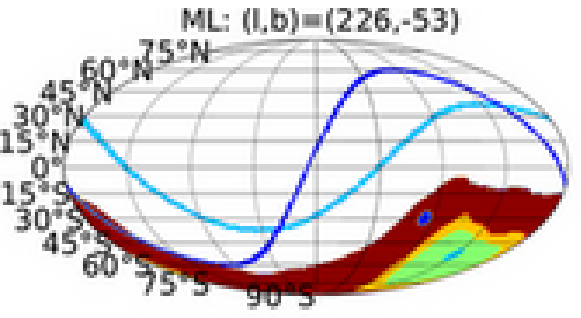}&
\includegraphics[width=0.166\textwidth]{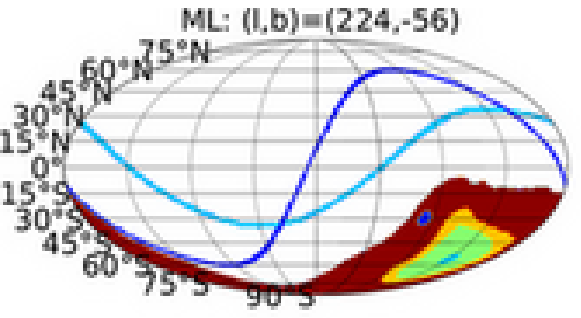}&
\includegraphics[width=0.166\textwidth]{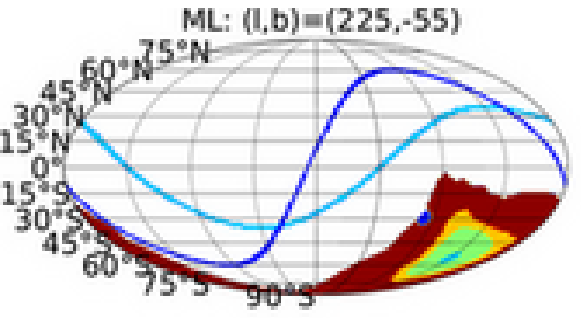}&
\includegraphics[width=0.166\textwidth]{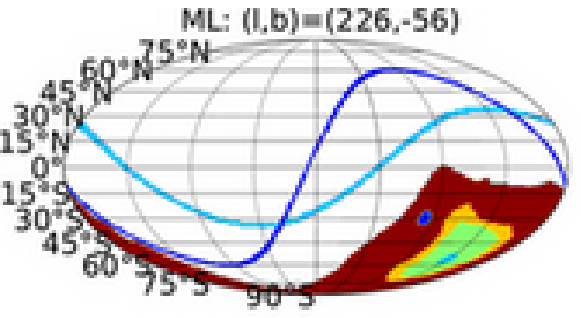}&
\includegraphics[width=0.166\textwidth]{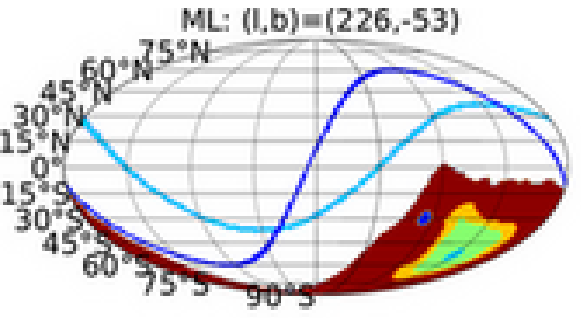}\\
$\lmin=6$ &
\includegraphics[width=0.166\textwidth]{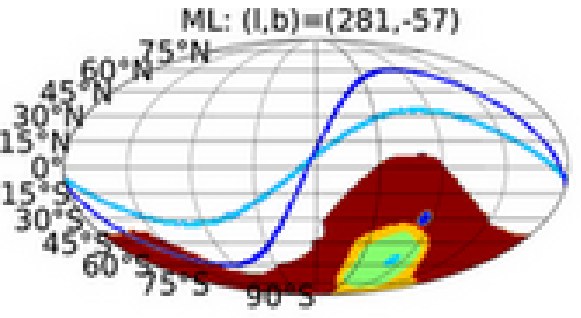}&
\includegraphics[width=0.166\textwidth]{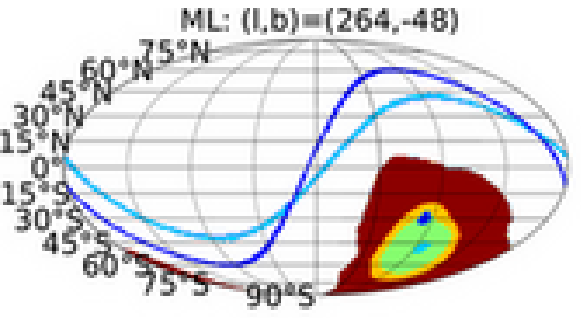}&
\includegraphics[width=0.166\textwidth]{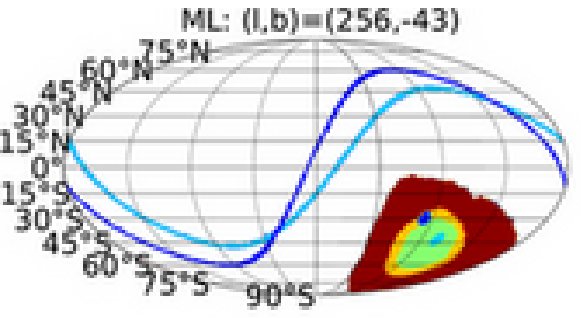}&
\includegraphics[width=0.166\textwidth]{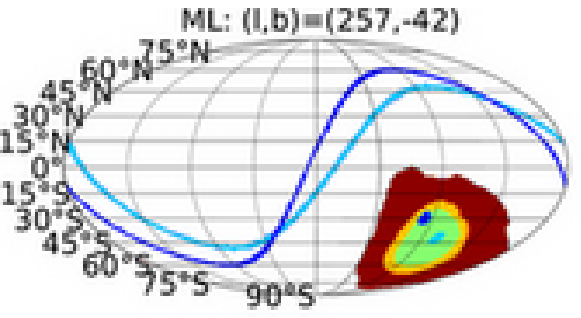}&
\includegraphics[width=0.166\textwidth]{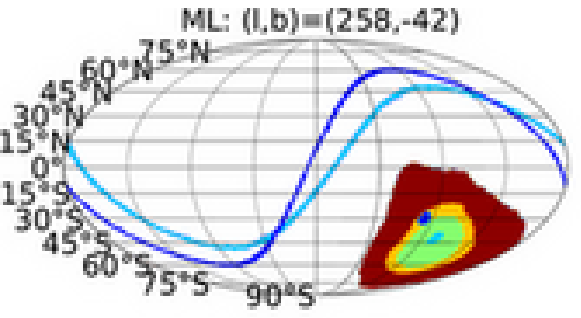}\\
&
$\lmin=19$ &
\includegraphics[width=0.166\textwidth]{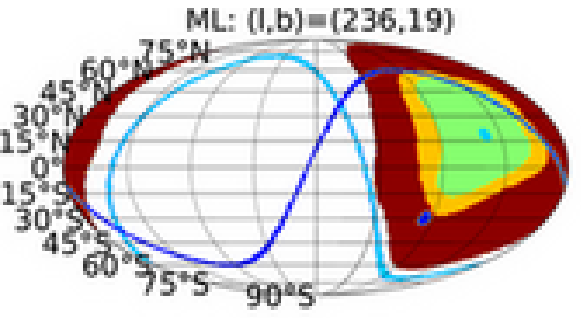}&
\includegraphics[width=0.166\textwidth]{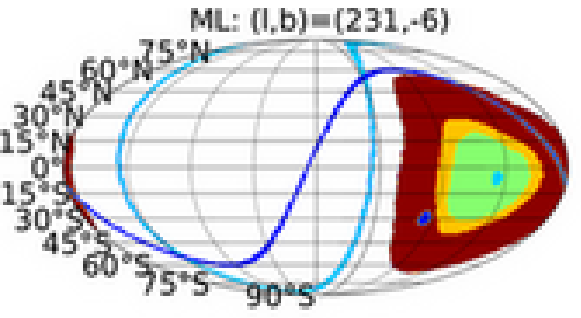}&
\includegraphics[width=0.166\textwidth]{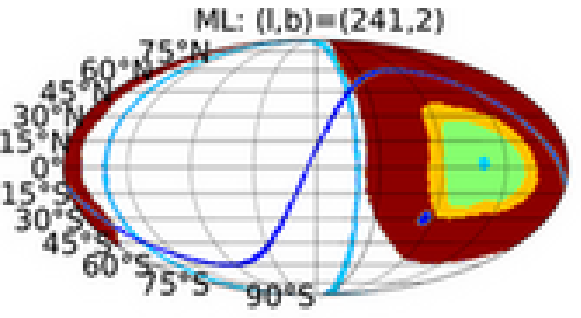}&
\includegraphics[width=0.166\textwidth]{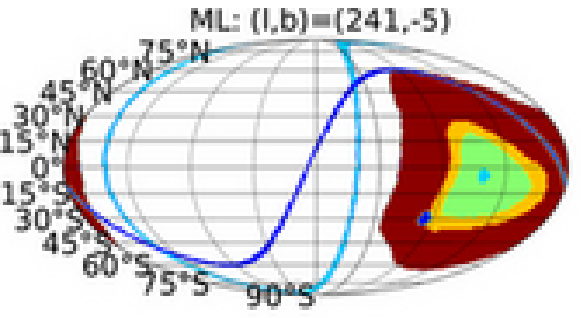}\\
&
&
$\lmin=29$ &
\includegraphics[width=0.166\textwidth]{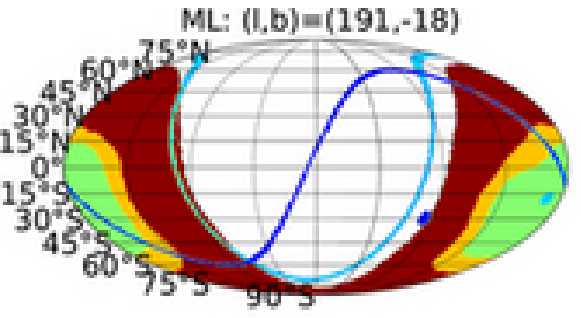}&
\includegraphics[width=0.166\textwidth]{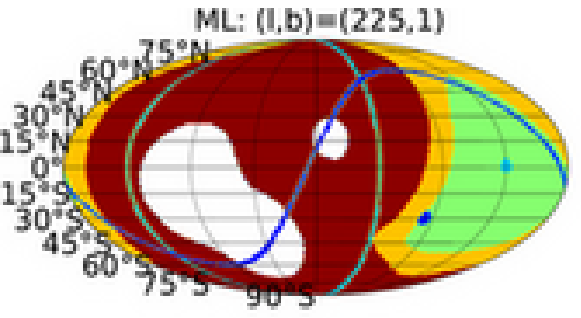}&
\includegraphics[width=0.166\textwidth]{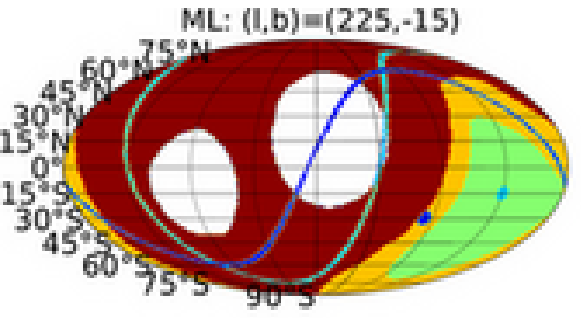}\\
&
&
&
$\lmin=39$ &
\includegraphics[width=0.166\textwidth]{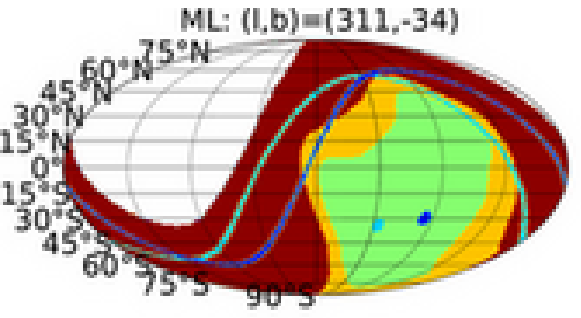}&
\includegraphics[width=0.166\textwidth]{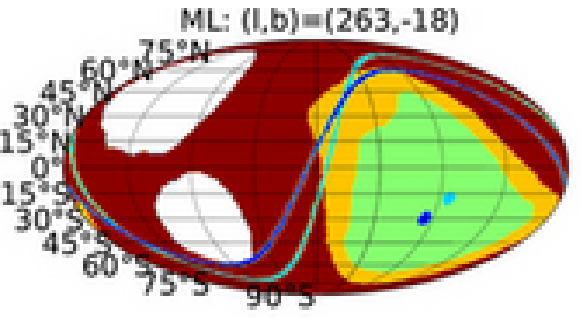}\\
&
&
&
&
$\lmin=59$ &
\includegraphics[width=0.16\textwidth]{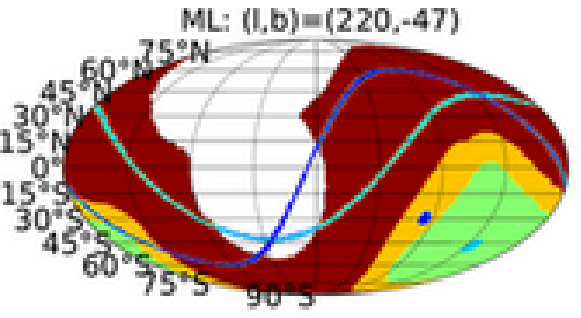}\\

\end{tabular}
\caption{Constraints on the modulation orientation from the V5 and ILC5 data.
For each considered multipole bin we plot the confidence regions corresponding to the $50\%$ (green), $68\%$ (yellow), 
and $95\%$ (red) confidence levels, based on the interpolated maps of the posterior distributions. 
In each map the maximum likelihood orientation and the corresponding dipole plane
are indicated using a light blue dot and line respectively. 
Additionally, for comparison, 
the ecliptic south pole and ecliptic plane are plotted in dark blue. 
In the top of each panel we give the galactic coordinates of the maximum likelihood solution. 
The arrangement of the panels is consistent with the cells in tables~\ref{tab:multipole_range},~\ref{tab:V5ILC5_ACLr6895}, and~\ref{tab:V5ILC5_CLrA0}.
}
\label{fig:V5ILC5_nmPDF}
\end{figure}

The best-fit orientations found in the analysis depend in general on the considered range of the multipoles. 
In particular, we see that the hemispherical power asymmetry, as measured here by the modulation orientations, 
generally tend to shift from larger galactic latitudes to the smaller galactic latitudes, as data of higher 
multipole bins are processed or cumulatively added up as $\lmax$ value increases. 
This was previously also noticed in \cite{Lew2008} but using a different method.
This effect seems to be seen in the first and second rows (as moving from the left to the right) for each of the 
data in table~\ref{tab:V5ILC5_ML}, 
or by comparing vertically in columns the first three rows as the low multipoles are removed from the analysis.
This is also seen directly in the distributions plotted in Fig.~\ref{fig:V5ILC5_nmPDF}, where 
it can be easily deduced in which bins
the effects of
small or vanishing modulations can be ignored. 
However we note that the effect is not present in every bin, and at most of order of few tens of degrees, 
and therefore we refrain from making any far-reaching speculations based on it.

While analyzing the distributions in Fig.~\ref{fig:V5ILC5_nmPDF} note that, those of them that correspond to very small 
or vanishing modal values of modulation amplitudes, have a very extended confidence level contours that cover large 
fraction of the sphere. These naturally result from a very flat likelihood function, and therefore any inference 
based on these cases is largely speculative and irrelevant. This is however expected, because 
for a vanishing modulation amplitude, there is no information on its orientation either.

We note that it is possible there is some degree of correlation between the plots for a given row, resulting from an 
cumulative effect of adding higher multipole bins.  The possible changes to the resulting distribution 
will jointly depend on the modulation amplitude and orientation in the 
added bin, but also on the amount of variance carried by the that bin as specified in table~\ref{tab:multipole_range}.

In table~\ref{tab:V5ILC5_ML} we plot the directions, in galactic coordinates, of the maximum posterior values 
found in the modulation orientation analysis.
Out of curiosity we also provide the angular separation of these directions from the ecliptic south pole 
to check for any possible extra alignments.

\begin{table}[!h]
\caption{Results of the modulation orientation parameter estimation for the V5 and ILC5 data.
The table contains the galactic coordinates of the maximum posterior modulation orientation and (in square brackets) 
the relative angular distance to the south ecliptic pole at \lb{276.4}{-29.8}.
}
\centering
\begin{tabular}{c|cccccc}\hline\hline
\multicolumn{7}{c}{V5 data}\\
$\lmin \backslash \lmax$ & 7 & 20 & 30 & 40 & 60 & 80\\
2 & ($281^\circ,-20^\circ)$, $[11^\circ]$ & ($233^\circ,-54^\circ)$, $[39^\circ]$ & ($234^\circ,-46^\circ)$, $[37^\circ]$ & ($225^\circ,-47^\circ)$, $[43^\circ]$ & ($220^\circ,-42^\circ)$, $[46^\circ]$ & ($223^\circ,-34^\circ)$, $[45^\circ]$ \\
6 && ($278^\circ,-68^\circ)$, $[38^\circ]$ & ($242^\circ,-55^\circ)$, $[35^\circ]$ & ($213^\circ,-42^\circ)$, $[52^\circ]$ & ($225^\circ,-35^\circ)$, $[43^\circ]$ & ($224^\circ,-31^\circ)$, $[45^\circ]$ \\
19 &&& ($187^\circ,2^\circ)$, $[90^\circ]$ & ($205^\circ,-8^\circ)$, $[70^\circ]$ & ($213^\circ,0^\circ)$, $[67^\circ]$ & ($217^\circ,1^\circ)$, $[64^\circ]$ \\
29 &&&& ($168^\circ,-19^\circ)$, $[96^\circ]$ & ($193^\circ,1^\circ)$, $[85^\circ]$ & ($212^\circ,-2^\circ)$, $[67^\circ]$ \\
39 &&&&& ($202^\circ,8^\circ)$, $[81^\circ]$ & ($236^\circ,1^\circ)$, $[49^\circ]$ \\
59 &&&&&& ($265^\circ,12^\circ)$, $[43^\circ]$ \\
\multicolumn{7}{c}{ILC5 data}\\
$\lmin \backslash \lmax$ & 7 & 20 & 30 & 40 & 60 & 80\\
2 & ($205^\circ,-67^\circ)$, $[56^\circ]$ & ($226^\circ,-53^\circ)$, $[43^\circ]$ & ($224^\circ,-56^\circ)$, $[45^\circ]$ & ($225^\circ,-55^\circ)$, $[44^\circ]$ & ($226^\circ,-56^\circ)$, $[44^\circ]$ & ($226^\circ,-53^\circ)$, $[43^\circ]$ \\
6 && ($281^\circ,-57^\circ)$, $[27^\circ]$ & ($264^\circ,-48^\circ)$, $[21^\circ]$ & ($256^\circ,-43^\circ)$, $[21^\circ]$ & ($257^\circ,-42^\circ)$, $[20^\circ]$ & ($258^\circ,-42^\circ)$, $[19^\circ]$ \\
19 &&& ($236^\circ,19^\circ)$, $[62^\circ]$ & ($231^\circ,-6^\circ)$, $[49^\circ]$ & ($241^\circ,2^\circ)$, $[46^\circ]$ & ($241^\circ,-5^\circ)$, $[42^\circ]$ \\
29 &&&& ($191^\circ,-18^\circ)$, $[77^\circ]$ & ($225^\circ,1^\circ)$, $[58^\circ]$ & ($225^\circ,-15^\circ)$, $[49^\circ]$ \\
39 &&&&& ($311^\circ,-34^\circ)$, $[30^\circ]$ & ($263^\circ,-18^\circ)$, $[17^\circ]$ \\
59 &&&&&& ($220^\circ,-47^\circ)$, $[46^\circ]$ \\
\end{tabular}
\label{tab:V5ILC5_ML}
\end{table}

\subsection{Modulation significance }
\label{sec:significance}
In the previous section we have shown that,  for some multipole ranges, the reconstructed, marginalized probability 
distribution function of the modulation amplitude, excludes the vanishing modulation value ($A=0$) at a very high 
confidence level.
It is important to ask whether this result is really robust, and whether or not we should reject the standard 
isotropic model of the Universe, at least, at some of the scales: i.e. those corresponding to the distributions 
with the strongest modulation detections, and the highest non-zero modulation significances. 
In particular, at least three ranges are of most concern: $\ell\in[7,19]$, and $\ell\in[7,79]$ 
for which the $A=0$ can be excluded at $99.5\%$ and $99.4\%$ CL respectively,
using the KQ75 sky-cut V5 data, and where the modulation parameters are constrained to be within ranges
$(0.07) 0.14 < 0.21 < 0.26 (0.31)$ at $68\%$ ($95\%$) CL and 
$(0.02) 0.05 < 0.08 < 0.11 (0.13)$ at $68\%$ ($95\%$) CL respectively.
Also the aforementioned range $\ell\in[7,39]$,
for which the $A=0$ can be excluded at $99.9\%$ CL 
using the full-sky ILC5 data, and where the modulation parameter is constrained to be within range
$(0.06) 0.10 < 0.13 < 0.17 (0.20)$ at $68\%$ ($95\%$) CL.

What we have done in the previous sections, is that we have estimated the best-fit modulation parameters 
(amplitude and orientation) with respect to the 
average from large amount of GRF simulations. Using average from large number of simulations ensures that we 
compare the data to really isotropic distributions, as any deviations from the statistical isotropy, even those resulting 
from the cosmic variance, will be averaged out.
Although the measurements also quantified the allowed magnitude of deviation from the ideal isotropy, allowed within the 
cosmic variance, via the standard deviation in the $\chisq$ tests,
there were number of explicit, or implicit assumptions or simplifications made on the way, like for example the usage 
of the $\chisq$ distributions, or neglecting the cosmic covariance effects, or residual foregrounds to name few. 
\begin{figure}[!t]
\centering
V5 data: $\ell\in[7,19]$\\
\includegraphics[width=\textwidth]{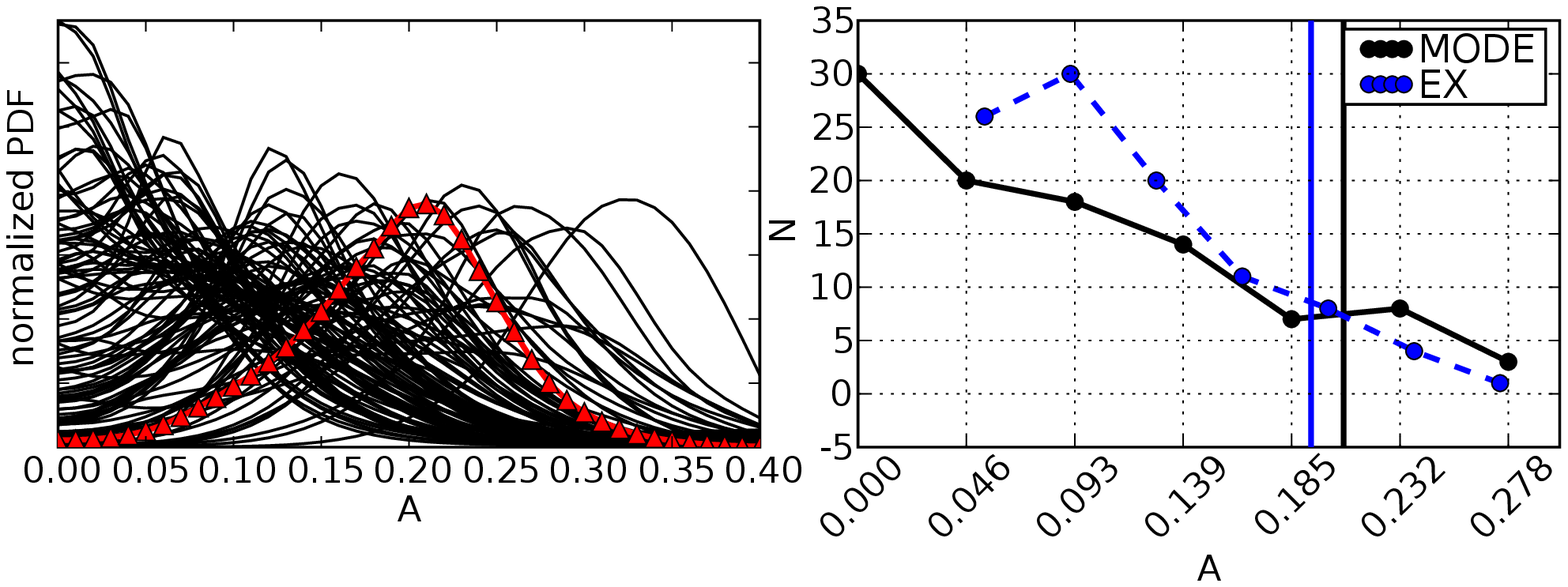}\\
V5 data: $\ell\in[7,79]$\\
\includegraphics[width=\textwidth]{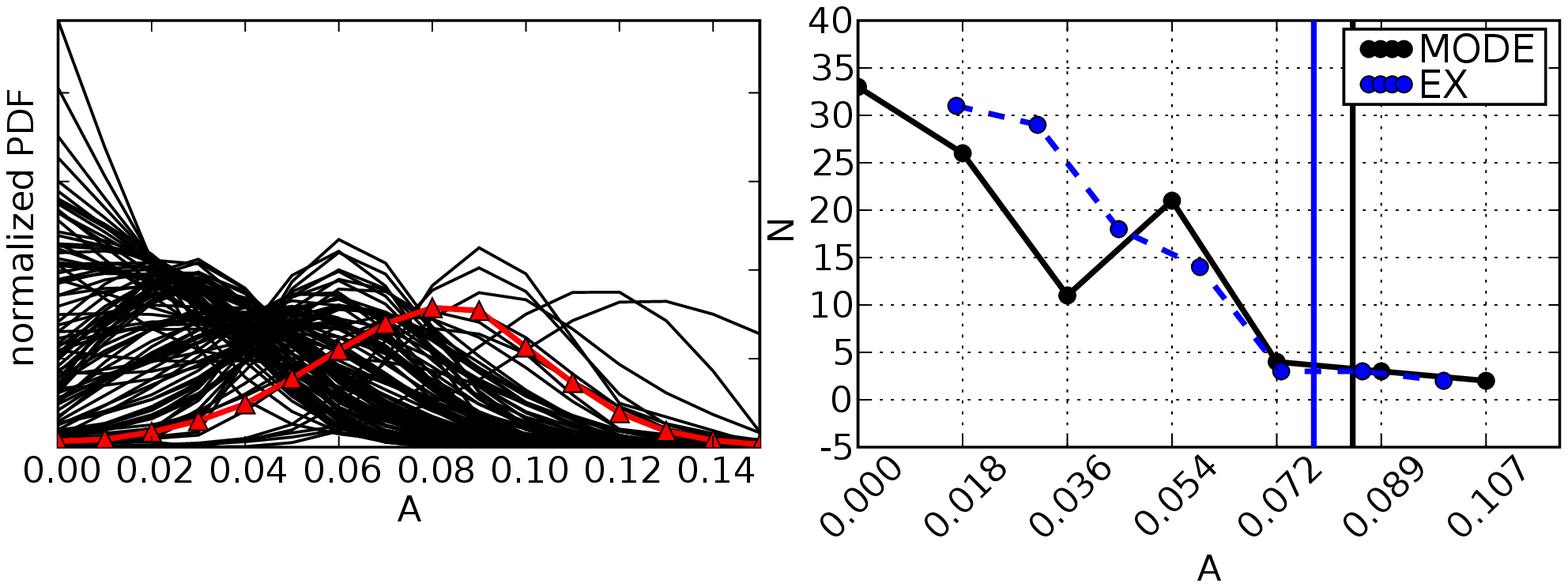}\\
ILC5 data: $\ell\in[7,39]$\\
\includegraphics[width=\textwidth]{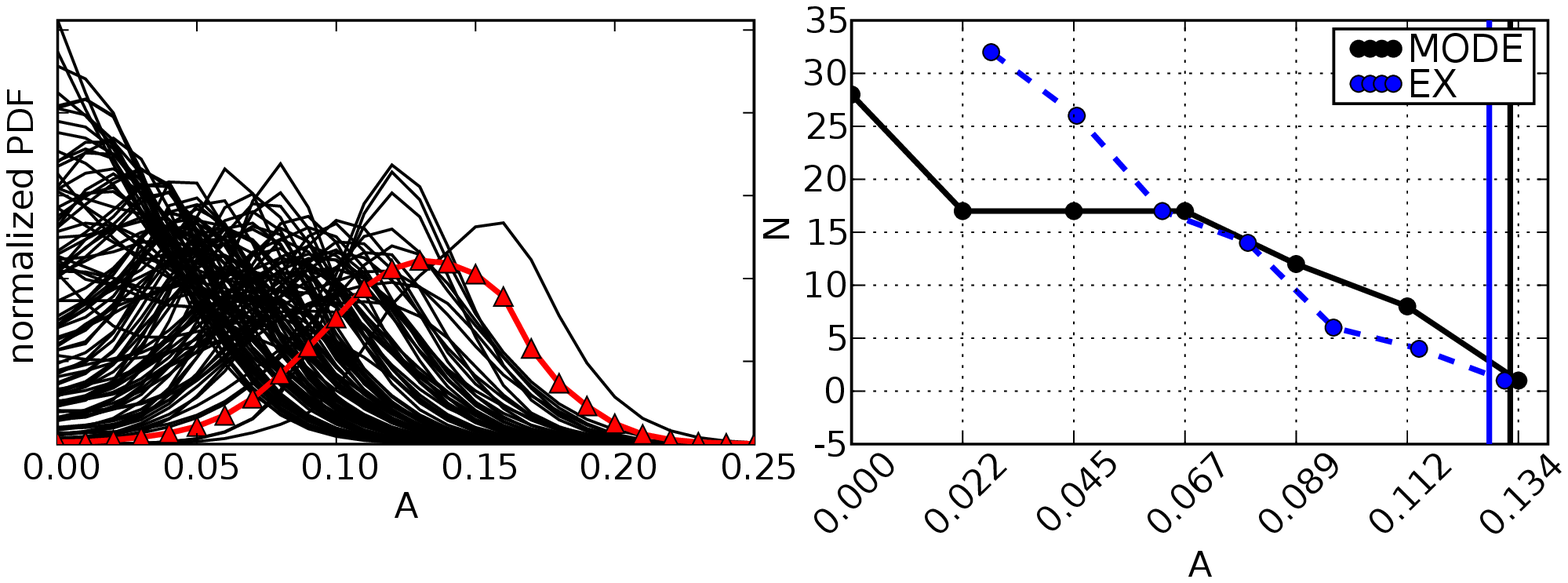}
\caption{In the left-hand side panels we plot a comparison of the reconstructed, marginalized distributions 
of the modulation amplitudes from 100 simulations of the WMAP V5 data (top and middle panels) 
and WMAP ILC5 data (bottom panel).
The WMAP data are plotted using red, thick lines (triangles). Only every 100th point of the interpolated, marginalized PDF is plotted.
In the right-hand side panels we plot the corresponding histograms of the expectancy values and modal values derived
from these distributions.
The WMAP data values are marked with vertical lines.
The plotted ranges yield the strongest, and most significant, plausible hemispherical anomalies in the data as 
inferred from the analysis in section~\ref{sec:modulation_amplitude}.
}
\label{fig:V5ILC5_modulation_significance}
\end{figure}

In \cite{Lew2008} we have performed the full covariance matrix analysis in two hemispherical regions with the same 
search parameter space as detailed in Section~\ref{sec:parameter_space}. We used 1000 GRF simulations 
(500 for covariance matrix estimation and 500 for probing the PDF of the underlying $\chisq$ distribution) 
and another 1000 simulations, modulated with amplitude of $A=0.114$.
The simulations were filtered up to $\lmax=40$. We found that on average about $8\%$ of the GRF simulations exhibited
a more unusual power distributions, than those found in the modulated simulations.
In this paper, although we use the same density in the search space, we have improved somewhat the method by using smooth 
interpolations.

In order to further test the robustness, and the significance of the power asymmetry anomalies, 
and circumvent all possible imperfections of the method, in the following we will pursue a similar test.
We process 100 GRF simulations of the V5 data through our parameter estimation pipeline, and compare the results
with the real data. Such approach should always be an ultimate test of the robustness, as it must give correct results
independently from the assumptions taken in the method. 

We will focus on the aforementioned multipole bins: $\ell\in[7,19]$ and $\ell\in[7,79]$ of the V5 data, 
and $\ell\in[7,39]$ of the ILC5 data. 
Within these ranges the power asymmetry seems to be very strong and very significant
(see. table~\ref{tab:V5ILC5_CLrA0} and figures~\ref{fig:V5_APDF} and~\ref{fig:ILC5_APDF}).

In Fig.~\ref{fig:V5ILC5_modulation_significance} we plot the results of the modulation parameter estimation for all 
tested simulations along with the WMAP data.
While it is clear that most of the simulations do not prefer any significant (if any at all) modulation amplitude values, 
at least few simulations, in our sample, yield modulations that are stronger in the considered 
range of multipoles, than those found in the data.
Also, from the shape of the PDF it is easy to infer that the significance of rejecting $A=0$ 
in these few cases will be even larger than in the case of the selected, most anomalous results from the WMAP data.

We find that 7 out of 100 V5 simulations yield stronger best-fit (modal) modulation amplitudes, 
and 6 of them also yield a more significant rejection of the $A=0$ parameter value, than the V5 data 
in the range $\ell\in[7,19]$. Similarly, for the range $\ell\in[7,79]$ 5 simulations yield stronger
and more significant best-fit modulation values.

Consequently, we conclude, that the significance, as inferred simply from integrating the PDF 
(as given in table~\ref{tab:V5ILC5_CLrA0}) is not quite robust. 
In light of these results we estimate the overall significance of possible modulation signals in the analyzed 
WMAP CMB maps at the level of about $\sim 94\%$ to $\sim 95\%$ depending on the particular range of multipoles. 
This remains greatly consistent with our previous results reported in~\cite{Lew2008} for the same data.

As for the ILC5 data we find that three simulations out of 100 yield a more significant rejection of $A=0$ hypothesis,
and curiously, only
one in those three also yields a stronger modulation within the multipole range $\ell\in[7,39]$.
Therefore, the corresponding overall significance of the power asymmetry, in this particular multipole range, is still
as high as about $\sim 99\%$. We note however, that since we did not use any sky masks in this case,
this result can probably be safely considered as somewhat overestimated, 
as any residual galactic foregrounds are likely only
to increase the level of the hemispherical power asymmetry, rather decrease it.

\section{Discussion}
\label{sec:discussion}

The results outlined in sections~\ref{sec:results} indicate that different multipole ranges
yield a different best fit modulation value, and that the modulation orientation
also  slightly varies from one multipole range to another.

In particular, the best-fit modulation orientation dependence when higher multipole bins are included cumulatively
is not as strong, as when the added multipole bins are considered individually.
Generally, we notice that within the best-fit orientations, that also yield a large modulation
values ($A\gtrsim 0.1$) the high-$\ell$ multipole bins prefer a close galactic plane orientation,
while the low-$\ell$ multipole bins rather prefer orientations with larger galactic latitudes.

The analysis of the modulation amplitude within few multipole bins yielded a large, best-fit modulation amplitudes,  
that seem to significantly reject the isotropic Universe model (with $A=0$).
However as much as few in one hundred GRF simulations, processed as data, also yielded a similar or larger modulation 
values, and also excluded the $A=0$ hypothesis at yet even higher confidence levels, than in the case of the V5 data. 
This effectively reduces the overall significance at which
the isotropic model of the Universe can be rejected, down to only about $94\%$ or $\sim95\%$ using the V5 data in the 
range $\ell\in[7,19]$, and $\ell\in[7,79]$ respectively.

We therefore pursued the analysis of the modulation signals in a two partially complementary ways. 
While the first approach addresses
the question of ``{\it how large and how significant is the best-fit modulation of the data?}'', the second approach
quantifies ``{\it how consistent is the best fit-modulation as compared with the GRF simulation expectations?}''.
The second approach should be more robust since it is free of any, possibly inaccurate, assumptions
that could result in underestimation of the size of the errors in the statistic, and in the result lead to spurious 
detections. These problems are effectively eliminated in a direct comparison with the GRF simulations.

Curiously the ILC5 data in the multipole range $\ell\in[7,39]$ still seem to be anomalous at a high
CL of about $99\%$; level almost as high as quoted in \citep{2007ApJ...660L..81E}.
However contrary to that work, we have not applied any sky masks to  this data,  and therefore
these results, given here only for comparison purposes with the V5 data, should still be 
treated with caution.

It would be interesting to perform similar analysis using the ILC5 data but with included  sky cut,
and to check the dependence of the analysis while varying the sky cuts from less to more aggressive.
Also, it could be interesting to check  the robustness and the significance in other multipole ranges 
than those two, tested in section~\ref{sec:significance}.
In principle, it would also be interesting to include other available renditions of the ILC maps, 
to see the stability of the modulation to different foregrounds cleaning pipelines.
We defer these issues for possible future work.

\section{Conclusions}
\label{sec:conclusions}

We performed tests of the hemispherical power asymmetry found in the CMB WMAP data for 
different bins of multipoles in two ways.

First, we introduced a statistic that searches for the orientation of two opposing, hemispherical regions
that maximize or minimize the hemispherical variance ratio, and compared these with the 
expectations from the GRF simulations. We found that the maximal asymmetry revealed this way is found
within a multipole range $\ell \in [8,15]$, with the southern hemisphere having larger variance
than the northern hemisphere. When these results are compared to the GRF simulations, the northern hemisphere
appears to be suppressed below the average expectation.

Secondly, we have introduced and tested a new method for measuring the power asymmetry in the CMB data,
as quantified within a bipolar modulation model \citep{2005PhRvD..72j3002G}. For the first time 
we constrained the modulation parameters as a function of various multipole  bins.
For each multipole range, we obtained the constraints on the 
the modulation amplitude and orientation. On the basis of the data sets, analyzed up to the maximal multipole
$\lmax=80$, i.e. the WMAP five-year inverse noise co-added, KQ75 sky cut map from the V channel (V5), 
and the five-year, full-sky, foregrounds cut ILC map (ILC5) we found that: \\
(i) generally the modulation amplitude decays as higher multipole bins are cumulatively added or independently analyzed,\\
(ii) the best fit modulation amplitude is small $A<0.03$ and insignificant for multipoles beyond $\ell\approx 40$\\
(iii) the most anomalous signals in terms of the modulation amplitude and its significance 
come from multipole range $\ell\in[7,19]$, and  $\ell\in[7,39]$ in the V5 and ILC5 data respectively.
For these ranges the significances of rejecting the isotropic cosmological model are $99.5\%$ and $99.9\%$ 
respectively and the constraints on the best fit, (PDF modal) modulation values are:
$(0.07) 0.14 < 0.21 < 0.26 (0.031)$ and
$(0.06) 0.10 < 0.13 < 0.17 (0.20)$ at $68\%$ ($95\%$) CL respectively.

Focusing on the two selected multipole ranges we performed an additional tests of the significance 
using GRF simulations processed as data, and found that similar or stronger and more significant
(in terms of rejecting the isotropic model) 
modulation values are obtained in 6 (1) cases in 100 simulations, which decreases the
overall significance of the power asymmetry in the CMB down to 94\% (99\%) in V5 (ILC5) data respectively.
To complement the results in the limit of high multipoles as well, we additionally tested the range 
$\ell\in[7,79]$ of the V5 data that also yields a strong and significant (99.4\%) best-fit modulation value
 - $(0.02) 0.05 < 0.08 < 0.11 (0.13)$ at $68\%$ ($95\%$) CL - 
but when this result was compared with the GRF simulations the effective significance is again decreased down 
to about 95\%.

Although the significance in case of the ILC5 data is still rather high, we warn that the results in this case 
were obtained without any sky cut, and therefore the asymmetry significance can be overestimated due to residual 
foregrounds.

Finally we note that a further analysis of the significance in terms of comparison with GRF simulations 
of other multipole ranges would be interesting, as well as analysis of the  power asymmetry in the ILC data
as a function of different sky cuts.

\section*{Acknowledegments}
BL would like to thank Naoshi Sugiyama for his support.
BL acknowledges the use of the computing facilities of the Nagoya University 
and support from the Monbukagakusho scholarship. 
\par We acknowledge the use of the Legacy Archive for
Microwave Background Data Analysis (LAMBDA). 
Support for LAMBDA is provided by the NASA Office of Space Science.  
\par We acknowledge the use of the NCAR's CSSGRID software for computing spline interpolations on sphere
and the GPL ccSHT package for scalar spherical harmonic transformations.

\bibliography{current} 
\bibliographystyle{aaeprint}

\end{document}